\documentclass[12pt,letter]{article}
\pdfoutput=1
\usepackage{graphicx, epsfig, color,cite}
\usepackage{amsmath}
\usepackage{amssymb}
\usepackage{caption,subcaption,graphicx}
\usepackage[hidelinks]{hyperref}
\def\eslt{\not\!\!\!{E_T}}
\def\to{\rightarrow}

\def\bi{\begin{itemize}}
\def\ei{\end{itemize}}
\def\te{\tilde e}

\def\tchi{\tilde\chi}

\def\tu{\tilde u}

\def\tb{\tilde b}
\def\tf{\tilde f}

\def\tst{\tilde t}
\def\ttau{\tilde \tau}

\def\tg{\tilde g}
\def\tnu{\tilde\nu}

\def\tw{\widetilde\chi^{\pm}}
\def\tz{\widetilde\chi^0}
\def\alt{\lesssim}
\def\agt{\gtrsim}
\def\be{\begin{equation}}  
\def\ee{\end{equation}}  
\def\bea{\begin{eqnarray}}  
\def\eea{\end{eqnarray}}

\begin{document}
\begin{titlepage}
\begin{flushright}
OU-HEP-191104
\end{flushright}

\vspace{0.5cm}
\begin{center}
{\Large \bf Mirage mediation from the landscape
}\\ 
\vspace{1.2cm} \renewcommand{\thefootnote}{\fnsymbol{footnote}}
{\large Howard Baer$^1$\footnote[1]{Email: baer@ou.edu },
Vernon Barger$^2$\footnote[2]{Email: barger@pheno.wisc.edu} and
Dibyashree Sengupta$^1$\footnote[3]{Email: Dibyashree.Sengupta-1@ou.edu}
}\\ 
\vspace{1.2cm} \renewcommand{\thefootnote}{\arabic{footnote}}
{\it 
$^1$Homer L. Dodge Department of Physics and Astronomy,
University of Oklahoma, Norman, OK 73019, USA \\[3pt]
}
{\it 
$^2$Department of Physics,
University of Wisconsin, Madison, WI 53706 USA \\[3pt]
}

\end{center}

\vspace{0.5cm}
\begin{abstract}
\noindent
Rather general considerations from the string theory landscape suggest
a statistical preference within the multiverse 
for soft SUSY breaking terms as large as possible subject to
a pocket universe value for the weak scale not greater than a 
factor of 2-5 from our measured value. 
Within the gravity/moduli-mediated SUSY breaking framework, 
the Higgs mass is pulled to $m_h\simeq 125$ GeV while 
first/second generation scalars are pulled to tens of TeV scale
and gauginos and third generation scalars remain at the few TeV range.
In this case, one then expects comparable moduli- and anomaly-mediated 
contributions to soft terms, leading to mirage mediation.
For an assumed stringy natural value of the SUSY $\mu$ parameter, 
we evaluate predicted sparticle mass spectra for mirage mediation 
from a statistical scan of the string landscape. 
We then expect a compressed spectrum of gauginos along with a 
higgsino-like LSP. 
For a linear (quadratic) statistical draw with gravitino mass 
$m_{3/2}\sim 20$ TeV, then the most probable mirage scale is 
predicted to be around $\mu_{mir}\sim 10^{13}$ ($10^{14}$) GeV.
SUSY should appear at high-luminosity LHC via higgsino pair 
production into soft dilepton pairs. Distinguishing mirage
mediation from models with unified gaugino masses may have to await
construction of an ILC with $\sqrt{s}>2m(higgsino)$.
\end{abstract}
\end{titlepage}

\section{Introduction}
\label{sec:intro}

So far, our only plausible understanding for the tiny, yet non-zero, value of 
the cosmological constant $\Lambda$ comes from Weinberg's 
multiverse explanation\cite{Weinberg:1987dv,Weinberg:1988cp}.
Assuming a vast array of pocket universes (PUs) within a broader 
multiverse\cite{Linde:2015edk},
each with different physical laws, then it may not be surprising that we
observe $\Lambda\sim 10^{-120} m_P^4$ (where $m_P$ is the reduced Planck mass
$m_P=2.4\times 10^{18}$ GeV) since if its value was much larger, 
then the cosmic expansion would be so fast that galaxies could not condense,
and observers would likely not arise.
This anthropic explanation depends on assuming a fertile patch of pocket 
universes within the multiverse which all have the Standard Model (SM)
as the low energy effective theory but for which the values of the cosmological
constant are spread uniformly across the decades of 
possible values\cite{Weinberg:2005fh}. 
Such reasoning allowed Weinberg to predict the value of $\Lambda$ to be within a
factor of a few, many years before its value was actually measured. 
These arguments were bolstered by the emergence of flux compactifications 
within string theory which provided the needed discretuum of meta-stable pocket
universes within the broader multiverse\cite{Bousso:2000xa,Susskind:2003kw,Ashok:2003gk}.

It is reasonable to ask if other scales, such as the weak scale, 
might arise from anthropic reasoning\cite{Agrawal:1997gf,Donoghue:2007zz}. 
In this case, it is usually assumed that the weak scale effective theory 
of the fertile patch is the
softly broken minimal supersymmetric standard model (MSSM) so that the weak 
scale is protected from large quantum mechanical corrections. 
In the context of string theory, typically a variety of hidden sectors
appear and several $F$- and $D$-term supersymmetry (SUSY) breaking 
vacuum expectation values (vevs) can contribute 
to the overall SUSY breaking scale\cite{Douglas:2004qg} 
which then in turn determines the weak scale via the scalar potential 
minimization conditions.
With a possibly non-minimal hidden sector, and with SUSY breaking vevs 
uniformly distributed across the decades of possible values, then statistically
{\it large} overall SUSY breaking scales are favored. 
Naively, one might expect as well that large values of the weak scale
would also be favored. Douglas has suggested a power law statistical
distribution for the overall soft SUSY breaking scale $m_{soft}$ of the form
$m_{soft}^n$ where $n=2n_F+n_D-1$, and where $n_F$ is the number of $F$- 
breaking fields and $n_D$ is the number of $D$-breaking fields 
contributing to the overall SUSY breaking scale\cite{Douglas:2004qg}. 

Several factors intervene to counteract this expectation\cite{Baer:2016lpj}. 
First, electroweak symmetry must be properly broken: no charge or color breaking
vacua (CCB) are allowed. 
Second, the soft terms must be such that $m_{H_u}^2$ is driven radiatively
to negative values so that EW symmetry is actually broken.
Third, the pocket universe value of the weak scale $m_{weak}^{PU}$ 
should be within a factor of a few from our measured value of the weak scale. 
Nuclear physics computations from Agrawal {\it et al.}\cite{Agrawal:1997gf} 
have shown that if $m_{weak}^{PU}\agt (2-5) m_{weak}^{OU}$ 
(OU stands for {\it our universe}) then
stable nucleons are all $\Delta^{++}$ baryons. 
Complex nuclei will not form and consequently atoms as we know them will not
form in such a universe. This anthropic requirement is known as the
{\it atomic principle} in that in order to have a universe with observers, 
then likely atoms (and consequently chemistry) as we understand them 
would have to be formed\cite{ArkaniHamed:2005yv}.

This picture has been explored in the context of gravity-mediated 
SUSY breaking models in the three-extra-parameter non-universal Higgs model
(NUHM3)\cite{Baer:2017uvn}. 
Motivated by the fact that all chiral matter superfields live in a
16-plet of $SO(10)$, the parameters of the NUHM3 model\cite{nuhm2} include
a common scalar mass for the first two generations $m_0(1,2)$
with a separate mass for the third generation $m_0(3)$. 
The Higgs superfields obtain independent soft terms $m_{H_u}^2$ and $m_{H_d}^2$
while the gauginos are unified at $m_{1/2}$ at the GUT scale. 
There are also common trilinear soft terms $A_0$ and bilinear $B$ although
$B$ is usually traded for the ratio of Higgs vevs $\tan\beta\equiv v_u/v_d$
via the scalar potential minimization conditions. 
It is also more convenient to trade the GUT scale values of 
$m_{H_u}^2$ and $m_{H_d}^2$ for the weak scale
values of the (SUSY conserving) Higgs/higgsino $\mu$ parameter and the
weak scale value of the pseudoscalar Higgs mass $m_A$. The superparticle mass
spectrum of the gravity-mediated NUHM3 model can then be calculated 
from the final form of the NUHM3 model parameter space:
\be
m_0(1,2),\ m_0(3),\ m_{1/2},\ A_0,\ \tan\beta,\ \mu,\ m_A\ \ \ \ (NUHM3) .
\ee

The approach of calculating SUSY particle mass spectra from the combined
statistical draw to large soft terms along with an anthropic veto of vacua
with improper EWSB or else too large a value of $m_{weak}^{PU}$ has met with
some considerable success. 
In Ref. \cite{Baer:2016lpj}, a qualitative examination was made and the 
flow of soft terms towards preferred 
statistical/anthropic values was shown to favor a large value of 
light Higgs mass $m_h\sim 125$ GeV while at the same time 
radiatively driving the {\it weak scale} soft terms towards natural values
comparable to the measured weak scale. In Ref. \cite{Baer:2017uvn}, a more 
quantitative approach generated probability distributions for 
Higgs and sparticle masses for the mild $n=1$ and 2 statistical 
draw to large soft terms. The Higgs mass probability histogram was found to peak at $m_h\simeq 125$ GeV which was understood in part because $A_0$ is pulled as
large as possible but stopping short of CCB minima- 
this lifts $m_h$ to $\sim 125$ GeV due to large mixing 
in the stop sector and hence large radiative corrections to $m_h$. 
Sparticle masses were typically pulled beyond LHC reach. 
In Ref. \cite{Baer:2019xww}, the SUSY landscape spectra from NUHM3 was 
confronted by various LHC sparticle and Higgs search limits and by 
WIMP dark matter search limits. 
Typical spectra would lie beyond both accelerator and dark matter search limits.
In Ref. \cite{Baer:2019uom}, in the context of SUSY axion models, it was 
shown that the draw to large PQ sector soft terms pulled one also to a large 
value of the PQ scale $f_a$. A large value of $f_a$ would generate too much 
axion dark matter {\it and} too much WIMP dark matter due to late time
axino and saxion decays in the early universe. It was concluded that
the PQ sector soft terms were likely correlated with MSSM soft terms
so that dark matter wouldn't be overproduced. 
In Ref. \cite{Baer:2019zfl}, it was shown that the draw to large first/second 
generation scalars could solve the SUSY flavor problem via a 
mixed decoupling/degeneracy solution since both first and second 
generation scalars would be drawn to large but common upper bounds in the 
20-30 TeV region.

In this paper, we extend this methodology to mixed gravity/moduli plus 
anomaly-mediated soft SUSY breaking (SSB) terms\cite{Choi:2005ge} 
in the context of the natural generalized mirage mediation model 
(nGMM)\cite{Baer:2016hfa}. 
Since the draw to large soft 
terms is related to a draw to large gravitino masses in supergravity,
then we would expect a gravitino mass $m_{3/2}$ in the tens of TeV regime
from SUSY on the landscape. But gaugino, third generation and Higgs soft terms
contribute to the weak scale either directly or via 1-loop
terms and so must instead lie in the TeV, not tens of TeV, regime.
In such circumstances, then one would expect comparable anomaly-mediated
and moduli-mediated (MM) contributions to soft terms-- a situation which 
requires mirage mediated rather then gravity-mediated only values for 
soft terms\cite{Choi:2005ge}. 

Our plan for this paper is then as follows. In the next 
Subsection \ref{ssec:brief}, we 
present a very brief review of some previous work on mirage-mediated 
SUSY breaking. Then, in Sec. \ref{sec:method} we present the
nGMM soft terms and parameter space and explain our methodology for
drawing the moduli-mediated soft terms to large values compared to the 
compulsory AMSB soft terms\cite{amsb} which only depend on the gravitino mass.
In Sec. \ref{sec:results}, we present histograms of probability for the
various nGMM parameters. 
These results include a prediction of the mirage scale
$\mu_{mir}$ where gaugino masses are expected to unify. 
The mirage scale prediction depends on the assumed value of $m_{3/2}$
and on the assumed power $n$ of the power-law selection 
of soft SUSY breaking terms from the landscape.
We also present probability histograms for the various
sparticle and Higgs masses. By measuring the gaugino masses directly
at colliders such as LHC or indirectly via splittings amongst the light
higgsino masses at a linear $e^+e^-$ collider\cite{Fujii:2017ekh}, 
then the mirage scale can be determined by running the weak scale 
gaugino masses up to the mirage scale\cite{Baer:2006tb}. 
Once the moduli/AMSB mixing parameter $\alpha$ is determined, 
then the associated gravitino mass $m_{3/2}$ can be determined. 
This in turn allows one to match against the predicted histograms for the mirage
scale for a particular value of $m_{3/2}$. 
We present a summary and conclusions in Sec. \ref{sec:conclude}.

\subsection{Brief review of some previous work on mirage mediation}
\label{ssec:brief}

The original mirage mediation scheme grew out of the 
Kachru-Kallosh-Linde-Trivedi (KKLT) 
proposal\cite{Kachru:2003aw} for moduli stabilization accompanied 
by some uplifting mechanism to gain a de Sitter minimum, 
{\it i.e.} a small cosmological constant from the landscape. 
The KKLT proposal was made in the context of IIB string 
theory compactified on an orientifold containing D3 and D7 branes. 
The complex structure or shape moduli and the dilaton could be stabilized 
by introducing NS and RR three-form fluxes with masses near the string scale.
A remaining single K\"ahler modulus $T$ would be stabilized by 
non-perturbative effects such as gaugino condensation or brane instantons, 
with $m_T\sim m_{3/2}\log (m_P/m_{3/2}) $, leading to a supersymmetric AdS vacuum. 
As a final step, an uplifting mechanism-- 
here the addition of an anti-D3 brane near the tip of a Klebanov-Strassler throat-- 
would raise the scalar potential of the theory to gain a de Sitter vacuum with
softly broken $N=1$ supersymmetry. 

In the KKLT scheme, a little hierarchy
\be
m_T\sim (4\pi^2) m_{3/2}\sim (4\pi^2)m_{soft}
\ee 
was expected to ensue\cite{Choi:2004sx,Choi:2005ge}, 
where $\log (m_P/m_{3/2})\sim 4\pi^2$ and where $m_{soft}$ is the expected 
scale of moduli (gravity)- mediated soft terms.
Since $m_{soft}$ was suppressed relative to $m_{3/2}$, then the 
moduli-mediated soft terms are expected to be comparable to contributions from
anomaly-mediation (which are suppressed relative to $m_{3/2}$ by
$\sim 1/(16\pi^2)$ loop factor). The resultant model has been dubbed
mirage-mediation\cite{LoaizaBrito:2005fa} (MM) due to the distinctive feature that 
gaugino (and scalar) masses evolve from non-universal values at the 
GUT scale to apparently universal values at some intermediate
scale 
\be
\mu_{mir}=m_{GUT}\cdot e^{(-8\pi^2/\alpha )}
\ee
where the introduced parameter $\alpha$ measures the relative 
moduli- versus anomaly-mediated contributions to gaugino 
masses\cite{Choi:2005uz,Falkowski:2005ck}.

Upon integrating out the heavy dilaton field and the
shape moduli, one is left with an effective broken supergravity theory
of the observable sector fields denoted by $\hat{Q}$ and the size
modulus field $\hat{T}$. The K\"ahler potential depends on the location
of matter and Higgs superfields in the extra dimensions via their
modular weights $n_i = 0 \ (1)$ for matter fields located on $D7$ ($D3$)
branes, or $n_i=1/2$ for chiral multiplets on brane intersections, while
the gauge kinetic function $f_a={\hat{T}}^{l_a}$, where $a$ labels the
gauge group, is determined by the corresponding location of the gauge
supermultiplets, since the power $l_a= 1 \ (0)$ for gauge fields on $D7$
($D3$) branes~\cite{Choi:2005uz,Falkowski:2005ck}.

Within the MM model, the SSB gaugino mass parameters, trilinear SSB
parameters and sfermion mass parameters, all renormalized just below the
unification scale (taken to be $Q=m_{\rm GUT}$), are given by,
\begin{eqnarray}
M_a&=& M_s\left( l_a \alpha +b_a g_a^2\right),\label{eq:M}\\
A_{ijk}&=& M_s \left( -a_{ijk}\alpha +\gamma_i +\gamma_j +\gamma_k\right),
\label{eq:A}\\
m_i^2 &=& M_s^2\left( c_i\alpha^2 +4\alpha \xi_i -
\dot{\gamma}_i\right) ,\label{eq:m2}
\end{eqnarray}
where $M_s\equiv\frac{m_{3/2}}{16\pi^2}$,
$b_a$ are the gauge $\beta$ function coefficients for gauge group $a$ and 
$g_a$ are the corresponding gauge couplings. The coefficients that
appear in (\ref{eq:M})--(\ref{eq:m2}) are given by
$c_i =1-n_i$, $a_{ijk}=3-n_i-n_j-n_k$ and
$\xi_i=\sum_{j,k}a_{ijk}{y_{ijk}^2 \over 4} - \sum_a l_a g_a^2
C_2^a(f_i).$ 
Finally, $y_{ijk}$ are the superpotential Yukawa couplings,
$C_2^a$ is the quadratic Casimir for the a$^{th}$ gauge group
corresponding to the representation to which the sfermion $\tf_i$ belongs,
$\gamma_i$ is the anomalous dimension and
$\dot{\gamma}_i =8\pi^2\frac{\partial\gamma_i}{\partial \log\mu}$.
Expressions for the last two quantities involving the 
anomalous dimensions can be found in the Appendices of 
Ref's.~\cite{Falkowski:2005ck,Choi:2006xb}.

The MM model is then specified by the parameters
\begin{equation}
\ m_{3/2},\ \alpha ,\ \tan\beta ,\ sign(\mu ),\ n_i,\ l_a. 
\label{eq:par1}
\end{equation}
The mass scale for the SSB parameters is dictated by the gravitino mass
$m_{3/2}$. The phenomenological parameter $\alpha$, which could be of
either sign, determines the relative contributions of anomaly mediation
and gravity mediation to the soft terms, and is expected to be $|\alpha|
\sim {\cal O}(1)$. Grand unification implies matter particles within
the same GUT multiplet have common modular weights, and that the $l_a$
are universal. We will assume here that all $l_a=1$ and, for
simplicity, there is a common modular weight for all matter scalars
$c_m$ but we will allow for different modular weights $c_{H_u}$ and
$c_{H_d}$ for each of the two Higgs doublets of the MSSM. Such choices
for the scalar field modular weights are motivated for instance by
$SO(10)$ SUSY GUT models where the MSSM Higgs doublets may live in different
$\bf 10$-dimensional Higgs reps.

Various aspects of MM phenomenology have been examined in
Refs.~\cite{Choi:2005uz,Falkowski:2005ck,Endo:2005uy,kn,bptw}. 
Of recent importance is to confront the MM models for various modular 
weight choices with the LHC Higgs mass discovery and also sparticle mass constraints.
By scanning over MM models with different $n_m$ and $n_H$
modular weight choices, but requiring $m_h=125\pm 2$ GeV, 
then all models were found to be rather highly fine-tuned 
in the electroweak using the conservative $\Delta_{EW}$ measure of 
fine-tuning\cite{Baer:2014ica}. 

The electroweak fine-tuning parameter~\cite{ltr,rns}, $\Delta_{\rm EW}$,
is a measure of the degree of cancellation between various contributions
on the right-hand-side (RHS) in the well-known expression for the $Z$ mass:
\be \frac{m_Z^2}{2} = \frac{m_{H_d}^2 +
\Sigma_d^d -(m_{H_u}^2+\Sigma_u^u)\tan^2\beta}{\tan^2\beta -1} -\mu^2
\simeq  -m_{H_u}^2-\Sigma_u^u-\mu^2 
\label{eq:mzs}
\ee 
which results from the minimization of the Higgs potential in the MSSM.
Here, $\tan\beta =v_u/v_d$ is the ratio of Higgs field
vacuum-expectation-values and the $\Sigma_u^u$ and $\Sigma_d^d$
contain an assortment of radiative corrections, the largest of which
typically arise from the top squarks. Expressions for the $\Sigma_u^u$
and $\Sigma_d^d$ are given in the Appendix of Ref.~\cite{rns}. 
If the RHS terms in Eq.~(\ref{eq:mzs}) are individually
comparable to $m_Z^2/2$, then no unnatural fine-tunings are required to
generate $m_Z=91.2$ GeV. $\Delta_{\rm EW}$ is defined to be the largest
of these terms, scaled by $m_Z^2/2$. Clearly, low electroweak
fine-tuning requires that $\mu$ be close to $m_Z$ and that $m_{H_u}^2$
be radiatively driven to {\it small} negative values close to the weak scale. 
This scenario has been dubbed radiatively-driven natural supersymmetry 
or RNS~\cite{ltr,rns}.

While the various MM models with particular discrete modular weight choices 
seem inconsistent with LHC Higgs mass measurements and sparticle mass limits, 
many general features of mirage mediation models were found to occur in 
a variety of different string based models. First, while the SSB terms were 
calculated within KKLT inspired set-ups including a single K\"ahler modulus
$T$, realistic string compactifications typically contain ${\cal O}(10-100)$ 
K\"ahler moduli. Under more general (and more plausible)  compactifications,
then it is reasonable to expect the general MM pattern of soft terms to ensue, 
but where the discrete modular weight choices are replaced by continuous 
parameters. For this reason, in Ref. \cite{Baer:2016hfa} a {\it generalized}
mirage mediation model (GMM) was proposed with continuous rather than discrete 
parameter choices, as detailed in Sec. \ref{sec:method}. The continuous parameters could allow for large trilinear soft terms $A_0$ which are needed to lift
$m_h$ to $\sim 125$ GeV whilst reducing fine-tuning in the 
$\Sigma_u^u(\tst_{1,2})$ terms. Also, the increased flexibility of GMM
allowed for small $\mu\sim 100-300$ GeV as expected from naturalness.

In addition, a wide variety of models were found to contain features of MM, 
but with important differences in the scalar sector. Indeed, 
Choi and Nilles\cite{Choi:2007ka} emphasize that the MM pattern of 
gaugino masses is rather general in a wide class of string-motivated models 
whilst deviations in the scalar sector are to be expected.
Examples include heterotic compactifications with a partially 
sequestered uplifting sector\cite{Lowen:2008fm}, IIB theory
with large volume compactifications\cite{Conlon:2006us,Choi:2010gm} 
(where gauginos are expected to adopt 
the mirage pattern but scalar masses are expected $\sim m_{3/2}$)
and heterotic orbifold compactifications\cite{Buchmuller:2005jr,Buchmuller:2006ik,Lebedev:2006kn,Lebedev:2006tr}. 
These latter models exhibit the phenomena of {\it local} 
grand unification\cite{Buchmuller:2005sh}
wherein different orbifold locations exhibit different gauge symmetries. 
For instance, first and second generation matter superfields may lie at
orbifold fixed points and so occur in complete {\bf 16} dimensional 
$SO(10)$ representations  with SSB scalar masses of order $m_{3/2}$. 
In contrast, the gauginos, Higgs superfields and third generation matter 
live more in the bulk, thus occuring in split representations and with 
SSB masses of order $m_{3/2}/(4\pi^2)$ and where the gaugino masses 
are expected with the mirage form. 
Thus, the sparticle mass spectra is expected to reflect the geography
of fields on the particular compactification manifold; this scheme 
is expected to be a more general result than just that which arises 
from any particular orbifold which has been selected\cite{Nilles:2014owa}.
The expected collider and dark matter phenomenology of such models has
been exhibited in Ref. \cite{Baer:2017cck}.
In addition, expectations for SUSY from 11-d $M$-theory models compactified
on a manifold of $G_2$ holonomy predict scalar masses of order
$m_{3/2}\sim 50-100$ TeV but with suppressed gaugino masses\cite{kane}. 
In such a case, one also expects comparable anomaly- and moduli-mediated
contributions to soft SUSY breaking terms.

\section{Methodology}
\label{sec:method}

In our approach, we will adopt the form of soft SUSY breaking terms expected
from general mirage mediation\cite{Baer:2016hfa} with a parameter space 
given by 
\be 
\alpha,\ m_{3/2},\ c_m,\ c_{m3},\ a_3,\ c_{H_u},\ c_{H_d},\
\tan\beta \ \  \ \ (GMM), 
\ee 
where $a_3$ is short for $a_{Q_3H_uU_3}$ (appearing in Eq. \ref{eq:A})
and $c_m$, $c_{m3}$, $c_{H_u}$ and $c_{H_d}$ arise in Eq. \ref{eq:m2}.
Here, we adopt an independent value $c_m$ for the first two
matter-scalar generations whilst the parameter $c_{m3}$ applies to
third generation matter scalars. 
In the GMM model, the $a_{ijk}$ and $c_i$ are elevated from discrete
to continuous parameters in order to accommodate more general
string theories and more general compactification schemes.
The independent values of $c_{H_u}$ and $c_{H_d}$, which set the
moduli-mediated contribution to the soft Higgs mass-squared soft terms, may
conveniently be traded for weak scale values of $\mu$ and $m_A$ as is
done in the two-parameter non-universal Higgs model (NUHM2)\cite{nuhm2}:
\be
\alpha,\ m_{3/2},\ c_m,\ c_{m3},\ a_3,\ \tan\beta , \mu ,\ m_A
\ \ \ (GMM^\prime ). \label{eq:gmmp}  \ee
This procedure allows for more direct exploration of 
stringy natural SUSY parameter space where most landscape solutions 
require $\mu\sim 100-300$ GeV in anthropically-allowed pocket 
universes\cite{Baer:2019cae}.

Thus, our final formulae for the soft terms are given by
\begin{eqnarray}
M_a&=& \left( \alpha +b_a g_a^2\right)m_{3/2}/16\pi^2,\label{eq:Ma}\\
A_{\tau}&=& \left( -a_3\alpha +\gamma_{L_3} +\gamma_{H_d} +\gamma_{E_3}\right)m_{3/2}/16\pi^2,\\
A_{b}&=& \left( -a_3\alpha +\gamma_{Q_3} +\gamma_{H_d} +\gamma_{D_3}\right)m_{3/2}/16\pi^2,\\
A_{t}&=& \left( -a_3\alpha +\gamma_{Q_3} +\gamma_{H_u} +\gamma_{U_3}\right)m_{3/2}/16\pi^2,\\
m_i^2(1,2) &=& \left( c_m\alpha^2 +4\alpha \xi_i -\dot{\gamma}_i\right)
(m_{3/2}/16\pi^2)^2 ,\label{eq:mi2} \\
m_j^2(3) &=& \left( c_{m3}\alpha^2 +4\alpha \xi_j -\dot{\gamma}_j\right)
(m_{3/2}/16\pi^2)^2 ,\\
m_{H_u}^2 &=& \left( c_{H_u}\alpha^2 +4\alpha \xi_{H_u} -\dot{\gamma}_{H_u}\right)
(m_{3/2}/16\pi^2)^2 ,\\
m_{H_d}^2 &=& \left( c_{H_d}\alpha^2 +4\alpha \xi_{H_d} -\dot{\gamma}_{H_d}\right)
(m_{3/2}/16\pi^2)^2 ,\label{eq:MHd}
\end{eqnarray}
where, for a given value of $\alpha$ and $m_{3/2}$, the values of
$c_{H_u}$ and $c_{H_d}$ are adjusted so as to fulfill the input values
of $\mu$ and $m_A$. 
In the above expressions, the index $i$ runs over first/second 
generation MSSM scalars
$i=Q_{1,2},U_{1,2},D_{1,2},L_{1,2}$ and $E_{1,2}$ while $j$ runs overs
third generation scalars $j=Q_3,U_3,D_3,L_3$ and $E_3$. 
The natural GMM model has been incorporated into the event generator 
program Isajet 7.86\cite{isajet} which we use here for spectra generation.

Douglas has proposed that the distribution of multiverse vacua 
versus hidden sector mass scale $m_{hidden}$ with a
given value of the weak scale $m_{weak}$ is represented by\cite{Douglas:2004qg}
\be
dN_{vac}[m_{hidden}^2,m_{weak},\Lambda ]= f_{SUSY}(m_{hidden}^2)\cdot
f_{EWSB}\cdot f_{CC}\cdot dm_{hidden}^2
\ee
where the soft SUSY breaking terms are related to the hidden sector mass scale
as $m_{soft}\sim m_{hidden}^2/m_P\sim m_{3/2}$. 
With many hidden sectors possible in string theory, 
then $m_{hidden}^4=\Sigma_i|F_i|^2+\sum_\alpha D_{\alpha}^2$
for the various $F$ and $D$ terms contributing to the totality of the 
SUSY breaking scale. Douglas observed that with $\hat{\Lambda}=3e^K|W|^2$
being the norm of the superpotential, then the cosmological constant is
\be
\Lambda=\sum_i|F_i|^2+\sum_\alpha D_{\alpha}^2-\hat{\Lambda}.
\ee
Since the superpotential $W$ receives additive contributions from many 
sectors of the theory, both supersymmetric and non-supersymmetric-- 
then one expects a uniform distribution in $\hat{\Lambda}$ and hence 
scanning in this variable fixes the cosmological constant 
to an anthropic value independent of the SUSY breaking scale. 
Thus, we may expect $f_{CC}\sim \Lambda/m_{string}^4$.

In addition, the total SUSY breaking scale is given by the distance 
from the origin in the space of all SUSY breaking parameters-- and in a high 
dimensional space, most of the volume is near the boundary. 
With a uniform distribution of individual SUSY breaking parameters, then
one expects a {\it power law} draw towards large SUSY breaking scale:
\be
f_{SUSY}\sim (m_{soft})^{2n_F+n_D-1}
\ee
where $n_F$ is the number of $F$- term breaking fields and $n_D$ is the number
of $D$-term breaking fields, and the factor of 2 arises because the 
$F$-breaking fields are complex whilst the $D$-breaking fields are real. 
Using this rather general ansatz, then already SUSY breaking by a single $F$
term implies a {\it linear} statistical draw to large soft terms. 
For multiple SUSY breaking fields, then the draw to large soft terms 
is even stronger.

An initial suggestion\cite{ArkaniHamed:2004fb} for $f_{EWSB}$ 
was that $f_{EWSB}\sim (m_{weak}/m_{soft})^2$ which would reflect 
the overall trend of multiplying the distribution by a naturalness measure. 
While this ansatz favors soft terms not too far 
removed from the weak scale, it fails in a number of cases\cite{Baer:2017uvn}. 
For instance, if trilinear soft terms become too big, one is forced into 
charge-and/or-color breaking minima of the Higgs potential. 
Such vacua must be vetoed rather than penalized by a statistical factor. 
Also, if $m_{H_u}^2$ becomes too large, then EW symmetry doesn't even break-- 
again, such vacua must be vetoed. 

Here, we will follow the nuclear physics results of 
Agrawal {\it et al.}\cite{Agrawal:1997gf} who found that in 
various pocket universes within our fertile patch
of MSSM effective theories, the generated value of the weak scale 
must be within a factor 2-5 of our measured value, lest all nucleons
turn into $\Delta^{++}$ baryons in which case complex nuclei, and hence atoms, 
will no longer form. Thus, here we obey the {\it atomic principle}\cite{ArkaniHamed:2005yv}: 
that complex life as we know it requires the existence of atoms, 
and consequently chemistry. 

To accommodate different weak scale values $m_{weak}^{PU}$ in 
different pocket universes, we invert the usual usage of Eq. \ref{eq:mzs}. 
We assume a natural solution to the SUSY $\mu$ problem (such as the
hybrid CCK or hybrid SPM models presented in Ref. \cite{Baer:2018avn} 
which generate a gravity-safe $U(1)_{PQ}$ symmetry for solving the strong CP
problem at the same time as generating $R$-parity conservation)\footnote{For 
a review of twenty solutions to the SUSY $\mu$ problem, 
see Ref. \cite{Bae:2019dgg}.} with $\mu\sim 200$ GeV.
(If $\mu\gg m_{weak}$, then only a relatively tiny fraction of vacua lead
to $m_{weak}\sim 100$ GeV\cite{Baer:2019cae}.) Then, instead of fixing 
$m_Z$ at its measured value in our universe, we can calculate its
pocket universe value $m_Z^{PU}$ for a given set of soft terms. 
We will require a value of $m_Z^{PU}<4m_Z(our\ universe)$ in accord with
Agrawal {\it et al.} which then corresponds to a value of $\Delta_{EW}<30$.
In this case, even with appropriate EWSB, large $A_0$ and $m_{H_u}^2$
terms actually lead to smaller contributions to the weak scale rather 
than larger ones (until one is forced into CCB or no EWSB vacua: 
see Ref's \cite{Baer:2016lpj,Baer:2019cae}). 
Thus, for $f_{EWSB}$ we will adopt 
\be
f_{EWSB}=\Theta (30-\Delta_{EW}) 
\ee
where each scan point leads to a different value of $m_Z^{PU}$.

To begin our scan over GMM$^{\prime}$ parameter points, we proceed as follows.
\bi
\item We select a particular value of $m_{3/2}$ which then fixes the 
AMSB contributions to SSB terms.
\item We also fix $\mu =200$ GeV for a natural solution to the SUSY 
$\mu$ problem.
This then allows for arbitrary values of $m_Z^{PU}$ to be generated 
but disallows any possibility of fine-tuning $\mu$ to gain $m_Z^{OU}$.
\ei
Next, we will invoke Douglas' power law selection of moduli-mediated 
soft terms relative to AMSB contributions within the GMM model.
Thus, for an assumed value of $n=2n_F+n_D-1$, we will generate
\bi
\item $\alpha^n$ with $\alpha:3-25$, a power law statistical selection 
for moduli-mediated gaugino masses $M_a$, $(a=1-3)$ over the gauge groups.
\item $(a_3\alpha)^n$, a power-law statistical selection of moduli-mediated
$A$-terms, with $(a_3\alpha):3-75$,
\item $m_0(1,2)\sim m_{3/2}$ so that $c_m=(16\pi^2/\alpha )^2$ so that 
first/second generation scalars are set maximally at $m_{3/2}$,
\item $(\sqrt{c_{m3}\alpha^2})^n$ to gain a power-law statistical 
selection on third generation scalar masses $m_0(3)$, 
with $(\sqrt{c_{m3}\alpha^2}): 3-80$
\item a power-law statistical selection on $m_{H_d}^2$ via $m_A^n$
with $m_A:300-7000$ GeV.
\item a uniform selection on $\tan\beta: 3-40$.
\ei
Our first informative scan allows us to narrow the range 
of $\alpha$ and $\sqrt{c_{m3}\alpha^2}$ while expanding the range of 
$a_3\alpha$, $m_A$ and $\tan\beta$. 
Our second scan proceeds with
\bi
\item $\alpha^n$ with $\alpha:5-20$, a power law statistical selection 
for moduli-mediated gaugino masses $M_a$, $(a=1-3)$ over the gauge groups.
\item $(a_3\alpha)^n$, a power-law statistical selection of moduli-mediated
$A$-terms, with $(a_3\alpha):3-100$,
\item $m_0(1,2)\sim m_{3/2}$ so that $c_m=(16\pi^2/\alpha )^2$ so that 
first/second generation scalars are set maximally at $m_{3/2}$,
\item $(\sqrt{c_{m3}\alpha^2})^n$ to gain a power-law statistical 
selection on third generation scalar masses $m_0(3)$, 
with $(\sqrt{c_{m3}\alpha^2}): 30-60$
\item a power-law statistical selection on $m_{H_d}^2$ via $m_A^n$
with $m_A:300-10000$ GeV.
\item a uniform selection on $\tan\beta: 3-50$.
\ei
followed by a focused scan by generating 
\bi
\item $\alpha^n$ with $\alpha:5-20$, a power law statistical selection 
for moduli-mediated gaugino masses $M_a$, $(a=1-3)$ over the gauge groups.
\item $(a_3\alpha)^n$, a power-law statistical selection of moduli-mediated
$A$-terms, with $(a_3\alpha):3-75$,
\item $m_0(1,2)\sim m_{3/2}$ so that $c_m=(16\pi^2/\alpha )^2$ so that 
first/second generation scalars are set maximally at $m_{3/2}$,
\item $(\sqrt{c_{m3}\alpha^2})^n$ to gain a power-law statistical 
selection on third generation scalar masses $m_0(3)$, 
with $(\sqrt{c_{m3}\alpha^2}): 30-60$
\item a power-law statistical selection on $m_{H_d}^2$ via $m_A^n$
with $m_A:1000-7000$ GeV.
\item a uniform selection on $\tan\beta: 3-40$.
\ei
We adopt a uniform selection on $\tan\beta$ since
this parameter is not a soft term.
Note that with this procedure-- while arbitrarily large soft terms are
statistically favored-- in fact they are all bounded from above since
once they get too big, they will lead either to non-standard EW vacua 
or else too large a value of $m_Z^{PU}$. In this way, models such as
split SUSY or high scale SUSY would be ruled out since for a 
natural value of $\mu$, then they
would necessarily lead to $m_Z^{PU}\gg (2-5)m_Z^{OU}$.

\section{Results for mirage mediation from the multiverse}
\label{sec:results}

In the following figures, we scan the soft terms of the GMM$^{\prime}$ model
according to the power law $m_{soft}^n$ for $n=1$ and 2 with
a fixed gravitino mass $m_{3/2}=20$ TeV.
Proceeding with much higher values of $m_{3/2}\agt 25$ TeV
always results in too-large of contributions to the weak scale when we take
$m_0(1,2)\simeq m_{3/2}$ (see Fig. 10 of Ref. \cite{Baer:2017cck}). 
We keep $\mu$ fixed at 200 GeV according to a natural
solution to the SUSY $\mu$ problem. 
We also veto non-standard EW vacua while for vacua with 
appropriate EWSB we require $f_{EWSB}=\Theta (30-\Delta_{EW})$ 
which corresponds to $m_Z^{PU}\le 4 m_Z^{OU}$. 
This latter anthropic selection imposes an upper bound on 
most GMM$^{\prime}$ parameters and  sparticle masses which would 
otherwise increase without limit according to $f_{SUSY}$.

\subsection{Parameters}

\begin{figure}[!htbp]
\begin{center}
\includegraphics[height=0.42\textheight]{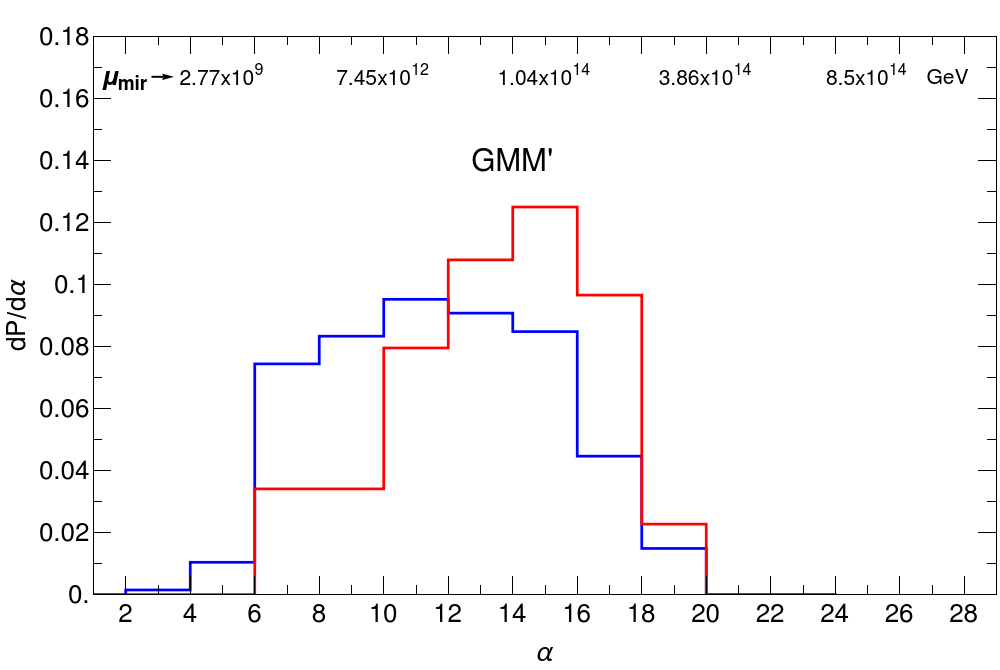}
\caption{Probability distribution for mixed moduli-anomaly mixing 
parameter $\alpha$ from $n=1$ (blue) and $n=2$ (red) statistical 
scans over the GMM$^{\prime}$ model with $m_{3/2}=20$ TeV. 
\label{fig:alpha}}
\end{center}
\end{figure}

In Fig. \ref{fig:alpha}, we first show the the normalized probability 
histogram $dP/d\alpha$ as a function of $\alpha$. 
The histogram is normalized to unit area. 
We also show for convenience on the upper scale various corresponding 
values of the gaugino mirage unification scale $\mu_{mir}$. 
From the figure, for a simple
linear draw ($n=1$ corresponding to SUSY breaking from a single $F$-term),
we see that the blue histogram has a rather broad peak spanning between
$\alpha\sim 6-16$ which then corresponds to a predicted mirage scale
$\mu_{mir}\sim 10^{10}-10^{14}$ GeV. There is relatively little probability
for $\mu_{mir}\alt 10^9$ Gev or for $\mu_{mir}\agt2\times 10^{14}$ GeV.
The mirage scale is actually testable in the GMM model since if we measure
any two of the three gaugino masses at the weak scale, then using
the known RGEs\cite{Martin:1993zk} we can extrapolate up in energy 
to see where they intersect. 
An intersection of all three gaugino masses at some intermediate mass scale 
would be strong supporting evidence for mirage mediation and would pick off
the requisite value of $\alpha$.

If instead we hypothesize an $n=2$ draw on soft terms, then we arrive at the 
red histogram. 
Here we see that the stronger statistical draw on moduli-mediated
soft terms results in a preference for higher $\alpha$ values peaked now 
at $\alpha\sim 15$ corresponding to $\mu_{mir}\sim 10^{14}$ GeV.
Substantial probability remains for $\mu_{mir}$ as low as $10^{11}$ GeV.

\begin{figure}[t]
  \centering
  {\includegraphics[width=.48\textwidth]{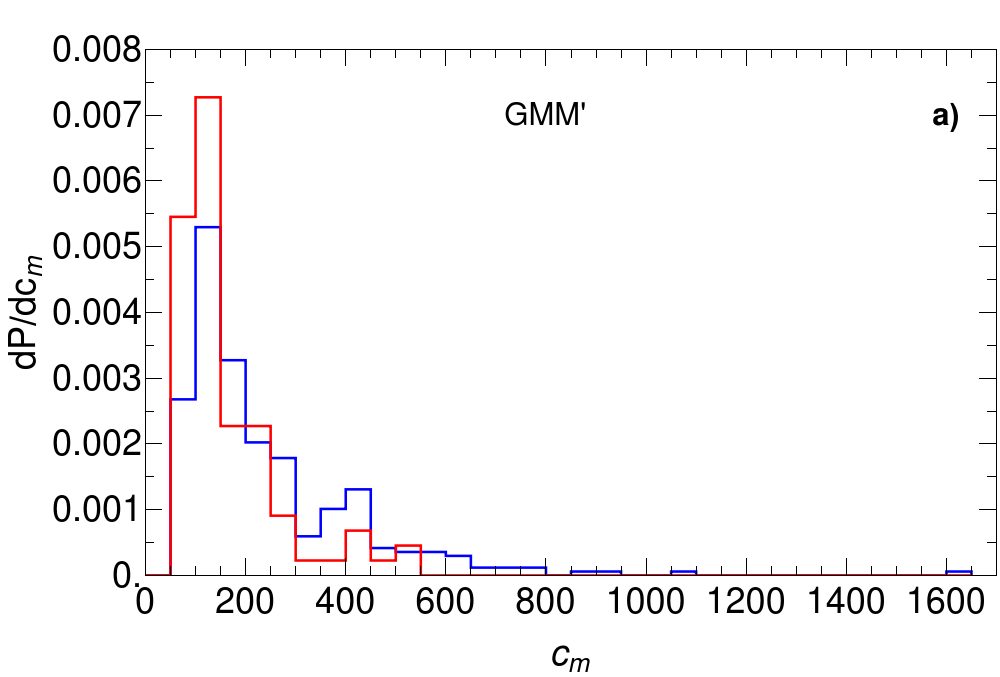}}\quad
  {\includegraphics[width=.48\textwidth]{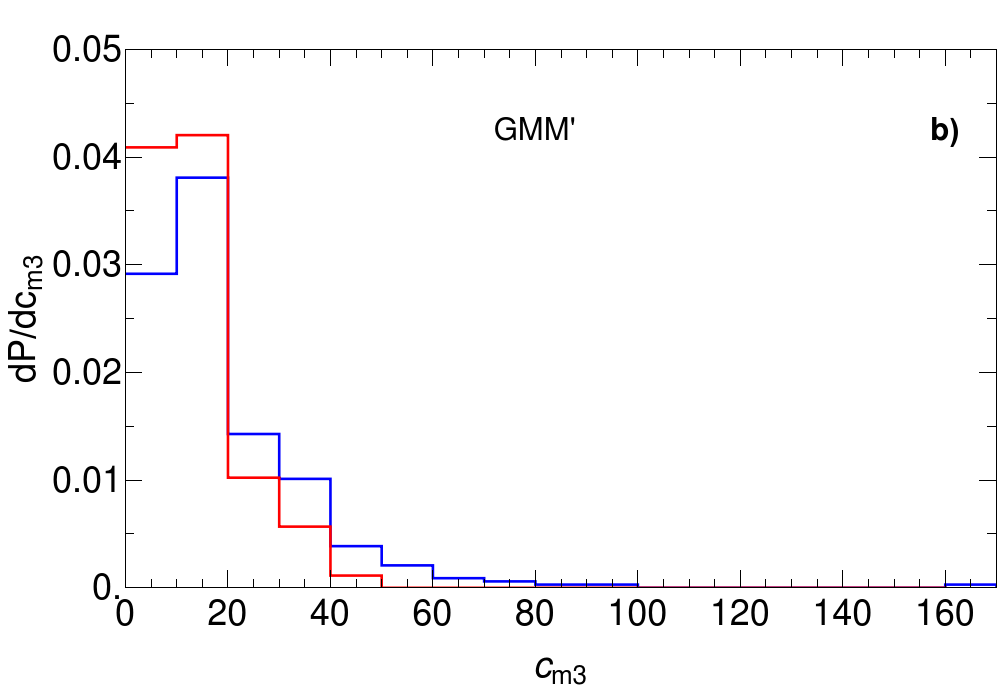}}\\ 
  {\includegraphics[width=.48\textwidth]{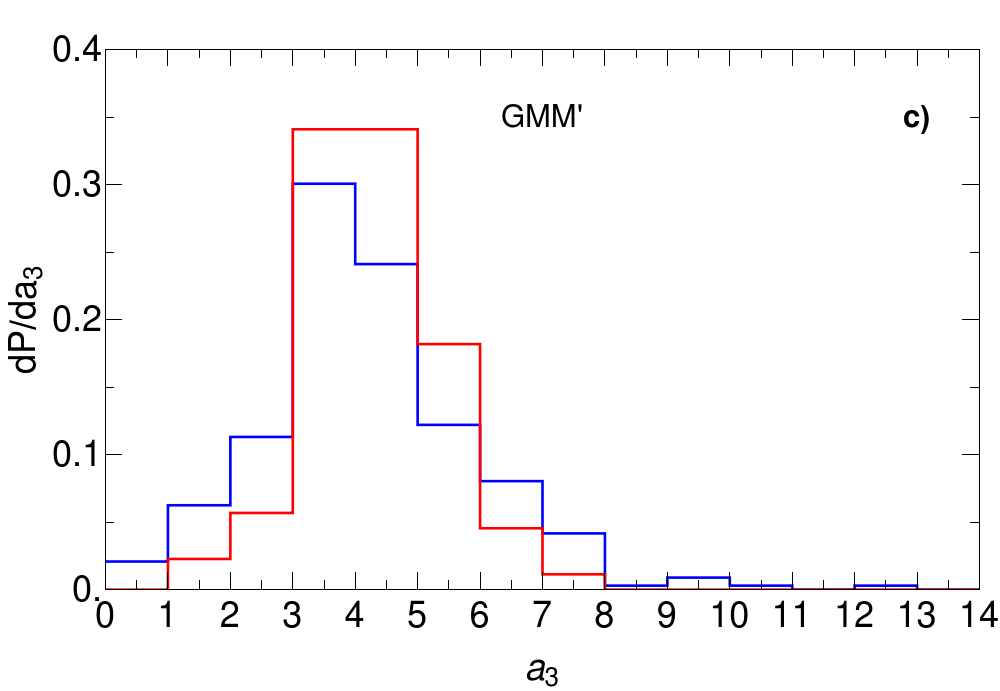}} \quad 
  {\includegraphics[width=.48\textwidth]{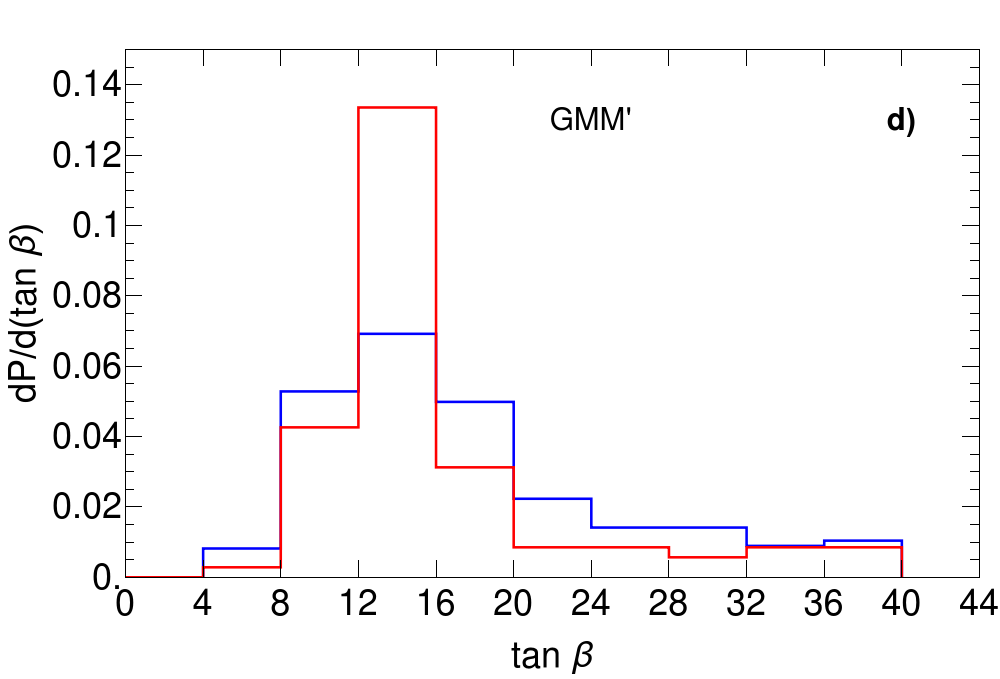}}\\
  \caption{Upper panels: Distributions in $c_m$ (left) and $c_{m3}$  (right). 
Lower panels: Distributions in $a_3$  (left) and $\tan\beta$ (right). 
Here, $n=1$ (blue) and $n=2$ (red) are from statistical scans 
over the nGMM model with $m_{3/2}=20$ TeV. 
}  
\label{fig:2}
\end{figure}

In Fig. \ref{fig:2}, we show histograms of probability for the other 
remaining parameters. In frame {\it a}), we show $dP/dc_m$ which peaks
for values of $c_m\sim 100-150$ for both $n=1$ and $n=2$. 
Since we have required $c_m=(16\pi^2/\alpha )^2$, this distribution just 
reflects the inverse-square distribution of $\alpha$ already shown in 
Fig. \ref{fig:alpha}. In frame {\it b}), we show the distribution in $c_{m3}$. 
In this case, we find values of $c_{m3}$ peaking at $c_{m3}\sim 5-15$
which sets the third generation matter scalar masses. 
These are more tightly restricted by the landscape since they largely 
determine the $\Sigma_u^u(\tst_{1,2})$ contributions to the weak scale. 
Since we cannot tune these away, then if they are too large we would have
$m_Z^{PU}\agt 4 m_Z^{OU}$ and we would violate the nuclear physics results 
of Ref. \cite{Agrawal:1997gf}.

In frame {\it c}), we show the distribution in $a_3$ which sets the 
magnitude of the moduli-mediated contribution to the trilinear soft term
$A_0$. Here, we find a statistical draw to large $-A_0$ terms with $a_3$
peaking around $3-6$. Such large $A_t$ terms actually {\it reduce } the
weak scale contributions $\Sigma_u^u(\tst_{1,2})$\cite{ltr,Baer:2019cae}.
At the same time, large $A_t$ terms yield maximal mixing in the stop sector
leading to an uplift of $m_h$ to $\sim 125$ GeV\cite{mhiggs,h125}.
If the $a_3$ parameter gets too big, then again large $\Sigma_u^u(\tst_{1,2})$ 
terms result while if even large values of $a_3$ occur then we are pushed
into CCB vacua (which must be vetoed). 

In frame {\it d}), we plot the distribution in $\tan\beta$, which was 
scanned uniformly. Here, we see the most probable value is 
$\tan\beta\sim 8-20$. For larger values of $\tan\beta\sim 20-50$, then the
$\tau$ and $b$-Yukawa couplings become large leading to large
$\Sigma_u^u(\tb_{1,2})$ contributions to the weak scale.
%
%

\subsection{Higgs and sparticle mass predictions}
\label{ssec:masses}

\begin{figure}[!htbp]
\begin{center}
\includegraphics[height=0.42\textheight]{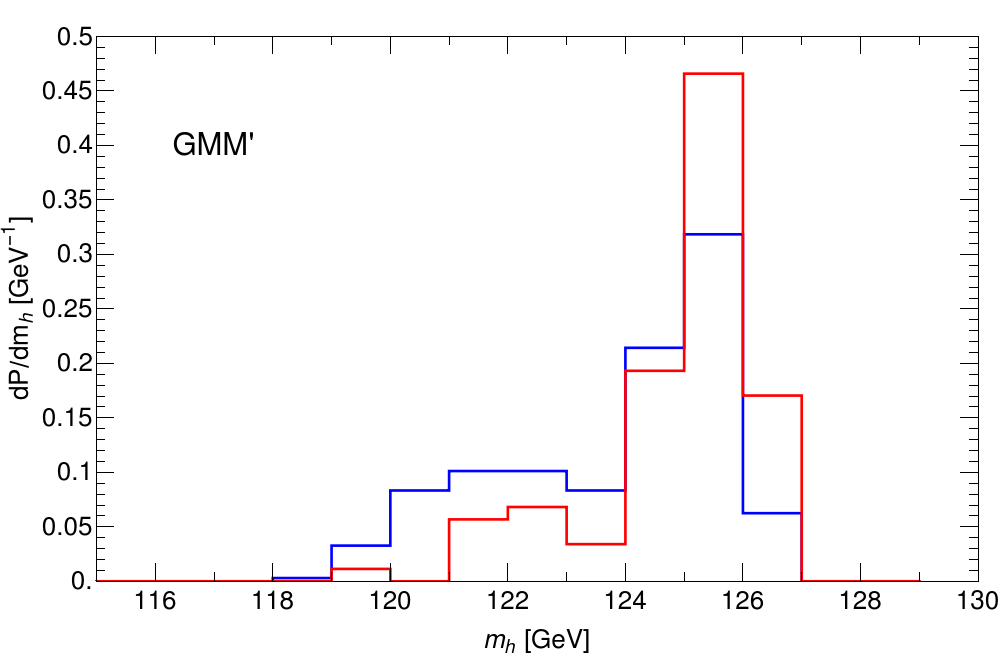}
\caption{Probability distribution for mass of light Higgs boson $m_h$
from $n=1$ (blue) and $n=2$ (red) statistical scans over nGMM model
with $m_{3/2}=20$ TeV.
\label{fig:mh}}
\end{center}
\end{figure}

In Fig. \ref{fig:mh}, we show the Higgs mass $m_h$ probability distribution
from the GMM model in the landscape for $m_{3/2}=20$ TeV with $n=1$ (blue) 
and $n=2$ (red). 
From the plot, we see that the most probable value of $m_h$ is 125 GeV 
for both cases. The value of $m_h$ reaches maximally 127 GeV 
but much higher values of $m_h$ always require $m_Z^{PU}>4 m_Z^{OU}$ from
the $\Sigma_u^u(\tst_{1,2})$ contributions to the weak scale.
These distributions are highly encouraging {\it post-dictions} of
the Higgs mass from general considerations of the string landscape!

\begin{figure}[t]
  \centering
  {\includegraphics[width=.48\textwidth]{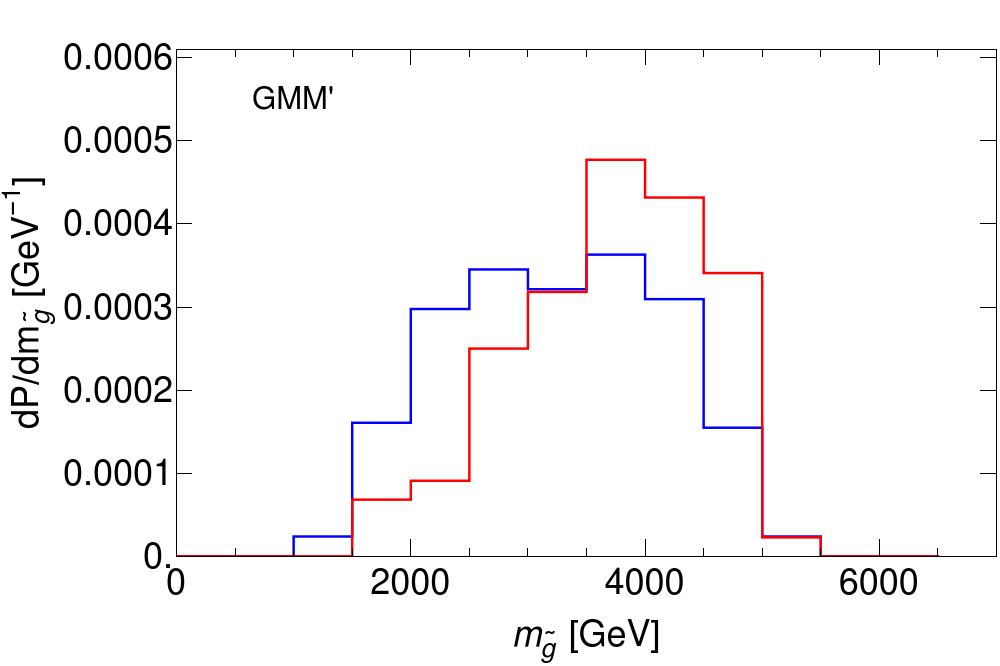}}\quad
  {\includegraphics[width=.48\textwidth]{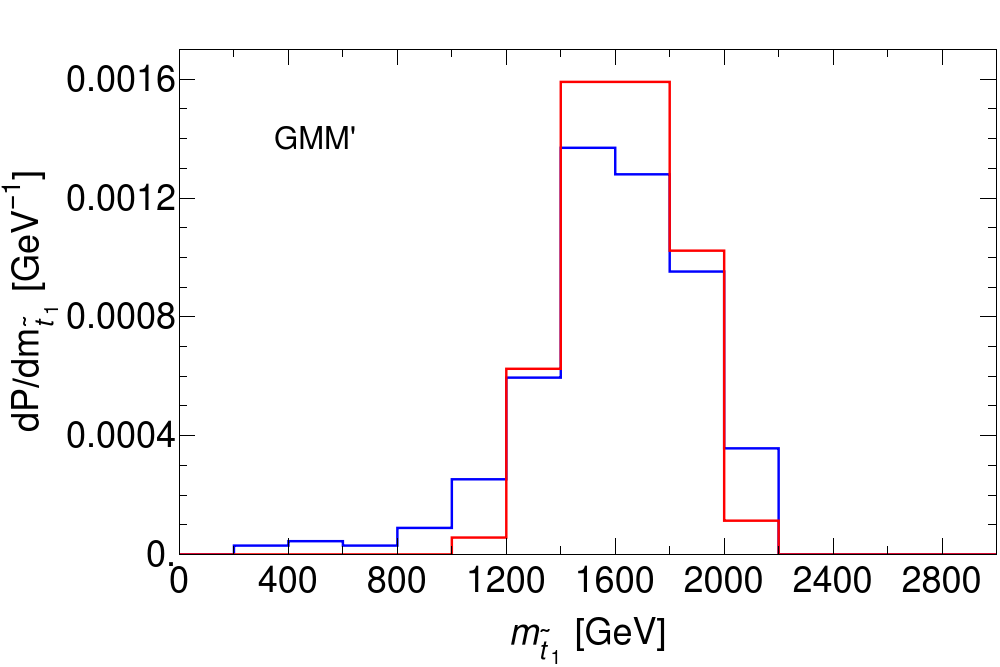}}\\ 
  {\includegraphics[width=.48\textwidth]{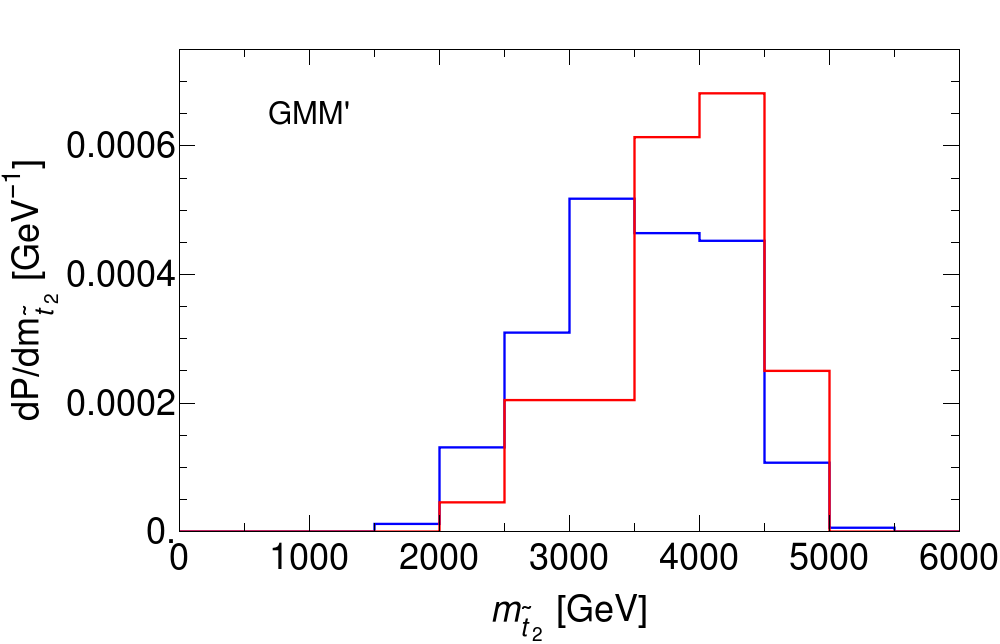}} \quad 
  {\includegraphics[width=.48\textwidth]{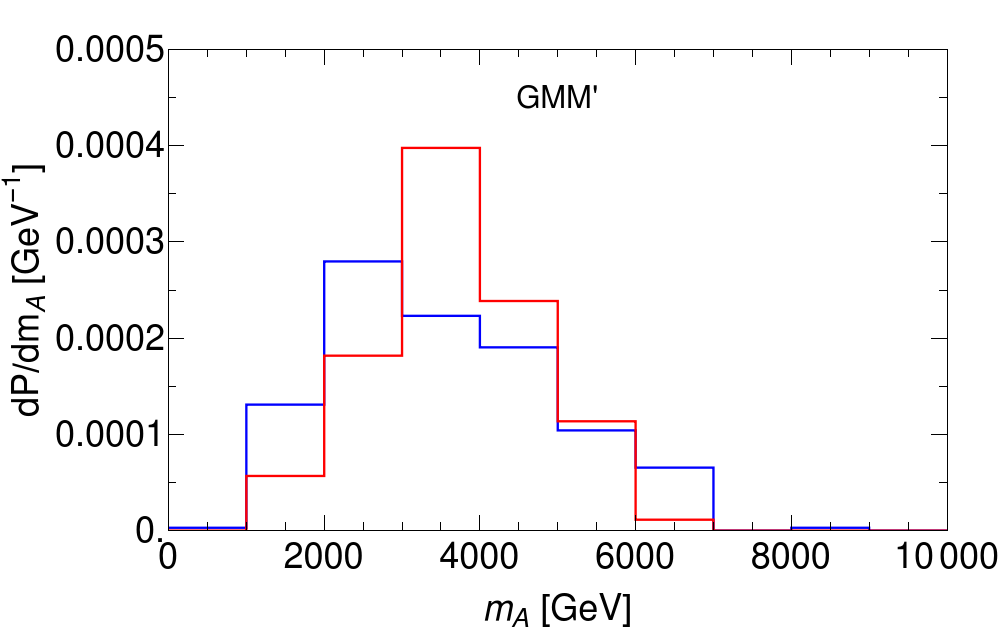}}\\
  \caption{Upper panels: Distributions in $m_{\tg}$ (left) and $m_{\tst_1}$  
(right). Lower panels: Distributions in $m_{\tst_2}$  (left) and 
$m_A$ (right). 
Here, $n=1$ (blue) and $n=2$ (red) are from a statistical scans 
over the nGMM model with $m_{3/2}=20$ TeV. 
}  
\label{fig:mass}
\end{figure}

In Fig. \ref{fig:mass}{\it a}), we show the probability distribution for
$m_{\tg}$ from the landscape within generalized mirage-mediation.
Here, we see that for $n=1$ with $m_{3/2}=20$ TeV, then $m_{\tg}\sim 2-5$ TeV, 
almost always safely beyond LHC Run 2 limits. 
For the $n=2$ case, then the distribution in $m_{tg}$ becomes somewhat 
harder with $m_{\tg}\sim 2.5-5$ TeV
with a most-probable value of $m_{\tg}\sim 4$ TeV. 
From these distributions, it seems reasonable that LHC has not yet discovered
SUSY via gluino pair production. 
The HL-LHC reach extends to $m_{\tg}\sim 2.7$ TeV\cite{Baer:2016wkz} 
while HE-LHC with $\sqrt{s}=27$ TeV will have a reach in $m_{\tg}$ 
to about 6 TeV\cite{Baer:2018hpb}.
Thus, discovery of SUSY via gluino pair production may have to await a
higher energy upgrade of LHC\cite{CidVidal:2018eel}.

In Fig. \ref{fig:mass}{\it b}), we show the probability distribution
for $m_{\tst_1}$. Here, we see for both $n=1$ and $n=2$ statistical draw, then
$m_{\tst_1}\sim 1-2$ TeV. These values of $m_{\tst_1}$ are generally
beyond current LHC top squark mass limits and so again it may be 
no surprise that LHC has not yet seen a signal via top-squark pair production.
While HL-LHC should have a reach in $m_{\tst_1}$ to about 1.5 TeV, 
the reach of HE-LHC extends to about 
$m_{\tst_1}\sim 3$ TeV\cite{Baer:2018hpb}. 
Thus, it may well require an energy upgrade of LHC to discover SUSY via 
top-squark pair production.

In Fig. \ref{fig:mass}{\it c}), we show the distribution in $m_{\tst_2}$.
In this case, we expect the landscape with GMM to yield a value
$m_{\tst_2}\sim 2.5-5$ TeV. 
Typically, we expect the higher range of these values to be beyond the 
reach of even HE-LHC.

In Fig. \ref{fig:mass}{\it d}), we show the expected probability for
the pseudoscalar Higgs mass $m_A$. We find that $m_A\sim 2-6$ TeV.
Such values are typically beyond the reach of HL-LHC\cite{Bae:2015nva}. 

\begin{figure}[t]
  \centering
  {\includegraphics[width=.48\textwidth]{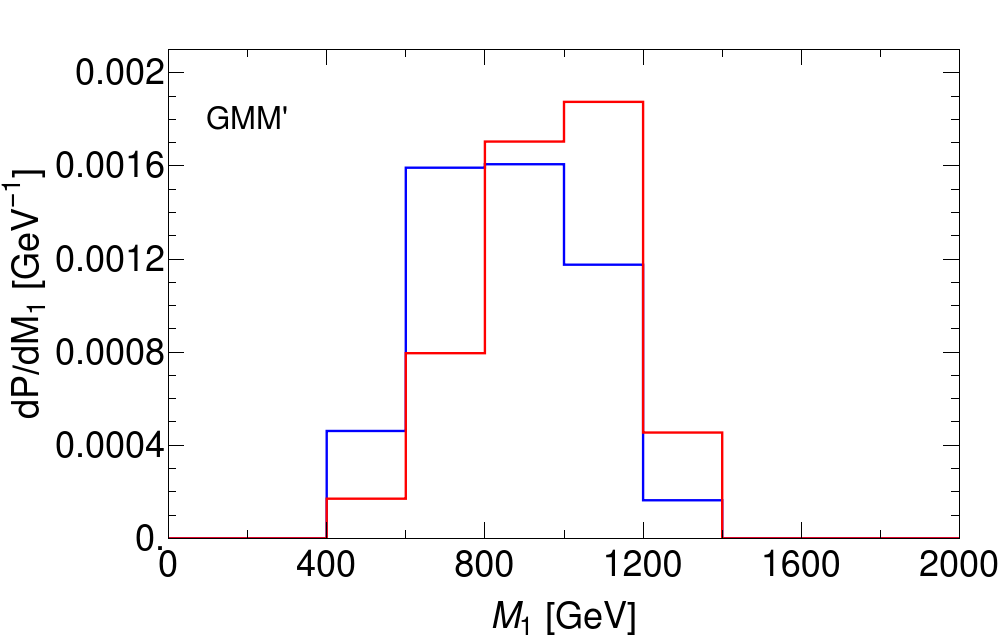}}\quad
  {\includegraphics[width=.48\textwidth]{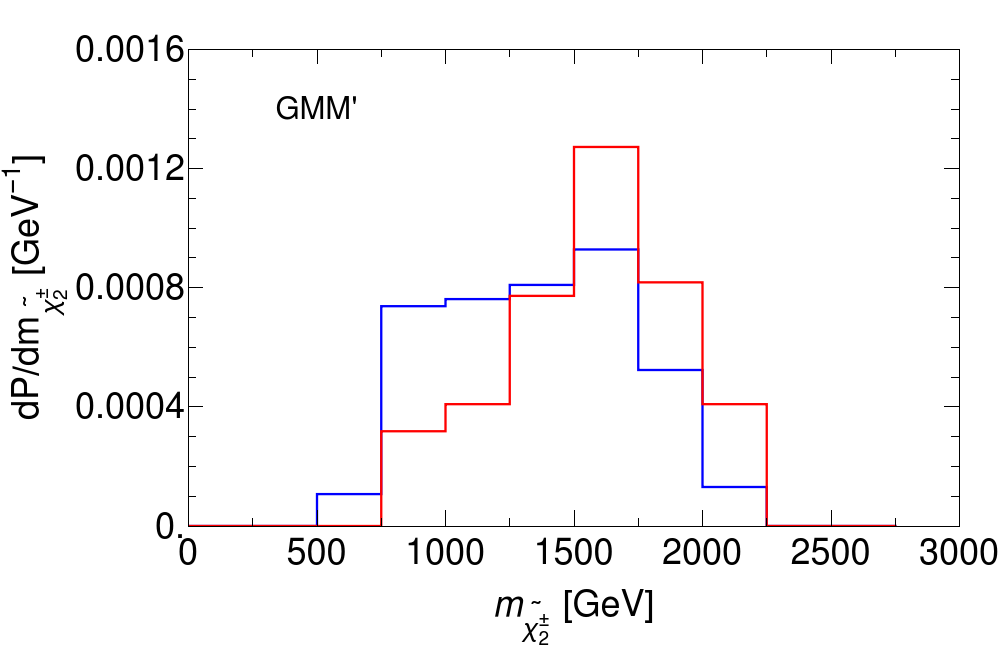}} 
  \caption{Distributions in $M_1$ (left) and $M_2$  
(right). 
Here, $n=1$ (blue) and $n=2$ (red) are from statistical scans over the nGMM model
with $m_{3/2}=20$ TeV. 
}  
\label{fig:Mi}
\end{figure}

One of the features of mirage-mediation is the expected compressed 
spectra of gauginos as compared to models with unified gaugino masses.
For unified gauginos, we expect weak scale gaugino masses in the ratio
$M_1:M_2:M_3\sim 1:2:6-7$. 
For the GMM model, these ratios can be quite different.
The $SU(3)$ gaugino mass $M_3\sim m_{\tg}$ 
(up to loop corrections) so that the approximate value of $M_3$ is given in
Fig. \ref{fig:mass}{\it a}). In Fig. \ref{fig:Mi}, we show the expected
electroweak gaugino masses. In frame {\it a}), the predicted bino mass
$M_1\sim 0.5-1.3$ TeV. This value is well above the expected value of
$\mu\sim 100-350$ GeV and so we would expect the lightest-SUSY-particle (LSP)
to be higgsino-like. The bino will be difficult to extract at LHC. However, 
a linear $e^+e^-$ collider with $\sqrt{s}>2m(higgsino)$ should be able
to pair produce higgsinos via reactions such as $e^+e^-\to \tchi_1^0\tchi_2^0$
 and measure the mass splitting $m_{\tchi_2^0}-m_{\tchi_1^0}$ 
which is sensitive to the bino mass\cite{Baer:2014yta}.
Such a machine should be able to extract $M_1$ to test the distribution 
in Fig. \ref{fig:Mi}{\it a}. 
In Fig. \ref{fig:Mi}{\it b}), we show the wino mass $M_2$ probability
distribution. 
It is expected that $M_2\sim 0.8-2.2$ TeV. 
The LHC can access wino pair production $\tchi_2^{\pm}\tchi_4^0$ via the
same-sign diboson signature\cite{Baer:2013yha,Baer:2017gzf} (SSdB) 
which is unique to SUSY models with light higgsinos:
$pp\to\tchi_2^{\pm}\tchi_4^0\to W^\pm W^\pm + \eslt$. 
The clean signature and signal production rate may 
allow one to extract a measurement of $M_2$ at HL- or HE-LHC via 
the total SSdB production rate.
Otherwise, again an $e^+e^-$ collider should be able to extract $M_2$ 
via the higgsino mass splittings which are measureable in higgsino pair
production reactions\cite{Baer:2014yta}.

\begin{figure}[t]
  \centering
  {\includegraphics[width=.48\textwidth]{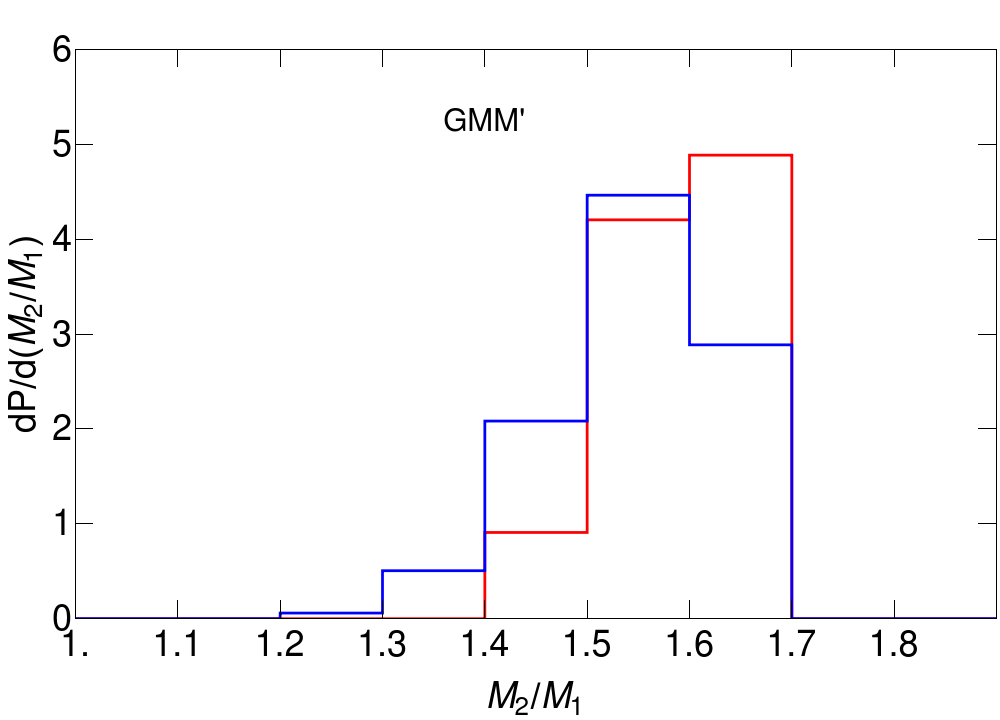}}\quad
  {\includegraphics[width=.48\textwidth]{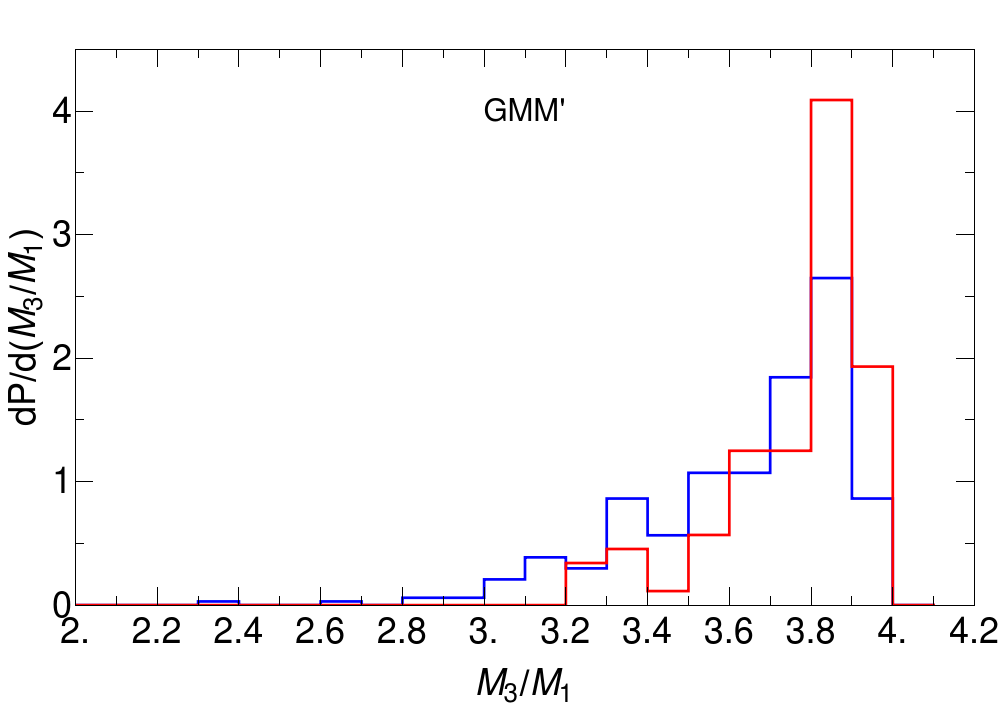}} 
  \caption{Distributions in $M_2/M_1$ (left) and $M_3/M_1$  
(right). 
Here, $n=1$ (blue) and $n=2$ (red) are from statistical scans over the nGMM model
with $m_{3/2}=20$ TeV. 
}  
\label{fig:MiRatio}
\end{figure}

In Fig. \ref{fig:MiRatio} we show the expected weak scale gaugino
mass ratios {\it a}) $M_2/M_1$ and {\it b}) $M_3/M_1$ 
which are expected from the landscape with mirage mediation. 
From frame {\it a}), we see that 
$M_2/M_1$ is expected to occur with ratio $\sim 1.4-1.7$ so that indeed the
electroweakinos are compressed, but not highly compressed. 
Such a compressed gaugino mass spectrum would be solid evidence for 
mirage-mediation\cite{Choi:2007ka}. In frame {\it b}), we find that
$M_3/M_1\sim 3-4$ rather than the expectation from gaugino-unified models 
where $M_3/M_1\sim 6-7$. 
While the gaugino mass spectrum is compressed, the
gap $m_{\tg}-m_{LSP}$ is actually greater than in gaugino-unified models since
the LSP is higgsino-like and close to the weak scale whilst gluinos 
are pulled statistically to large values.

\begin{figure}[!htbp]
\begin{center}
\includegraphics[height=0.42\textheight]{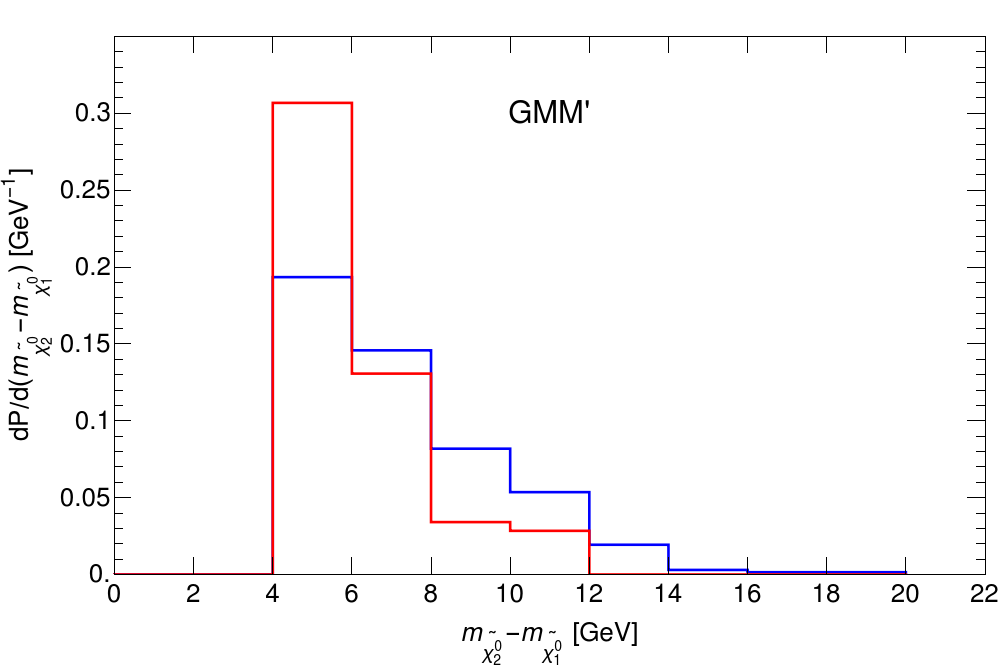}
\caption{Probability distribution for light neutral
higgsino mass difference $m_{\tz_2}-m_{\tz_1}$
from $n=1$ (blue) and $n=2$ (red) statistical scans over the nGMM model
with $m_{3/2}=20$ TeV.
\label{fig:mz2mz1}}
\end{center}
\end{figure}

We also plot in Fig. \ref{fig:mz2mz1} the expected $m_{\tchi_2^0}-m_{\tchi_1^0}$ 
mass gap. 
This gap is expected to be directly measurable at LHC via
the higgsino pair production reaction $pp\to\tchi_1^0\tchi_2^0$ 
followed by $\tchi_2^0\to \tchi_1^0 \ell^+\ell^-$\cite{SDLJMET}. 
(Indeed, there appears already some excess in this channel at Atlas with
139 fb$^{-1}$; see Fig. 10{\it a}) of Ref. \cite{ATLAS:2019lov}.) 
From the plot, we see the mass gap is typically 
$m_{\tchi_2^0}-m_{\tchi_1^0}\sim 4-12$ GeV so the opposite-sign (OS) dileptons 
will likely be quite soft.
This discovery channel for SUSY appears to be the {\it most propitious one} 
for HL-LHC\cite{Baer:2019xww}.

\subsection{$m_0^{MM}$ vs. $m_{1/2}^{MM}$ parameter space for $m_{3/2}=20$ TeV}
\label{ssec:m0vsmhf}

A panoramic view of some of our essential conclusions may be displayed in
the $m_0^{MM}$ vs. $m_{1/2}^{MM}$ plane which is then analogous to the
$m_0$ vs. $m_{1/2}$ plane of the mSUGRA/CMSSM or NUHM2,3 models.
Here, we define $m_0^{MM}=\sqrt{c_m}\alpha (m_{3/2}/16\pi^2)$ which is the
pure moduli-mediated contribution to scalar masses. The moduli-mediated
contribution to gaugino masses is correspondingly given by
$m_{1/2}^{MM}\equiv \alpha m_{3/2}/(16\pi^2)$.

\begin{figure}[t]
  \centering
  {\includegraphics[width=.48\textwidth]{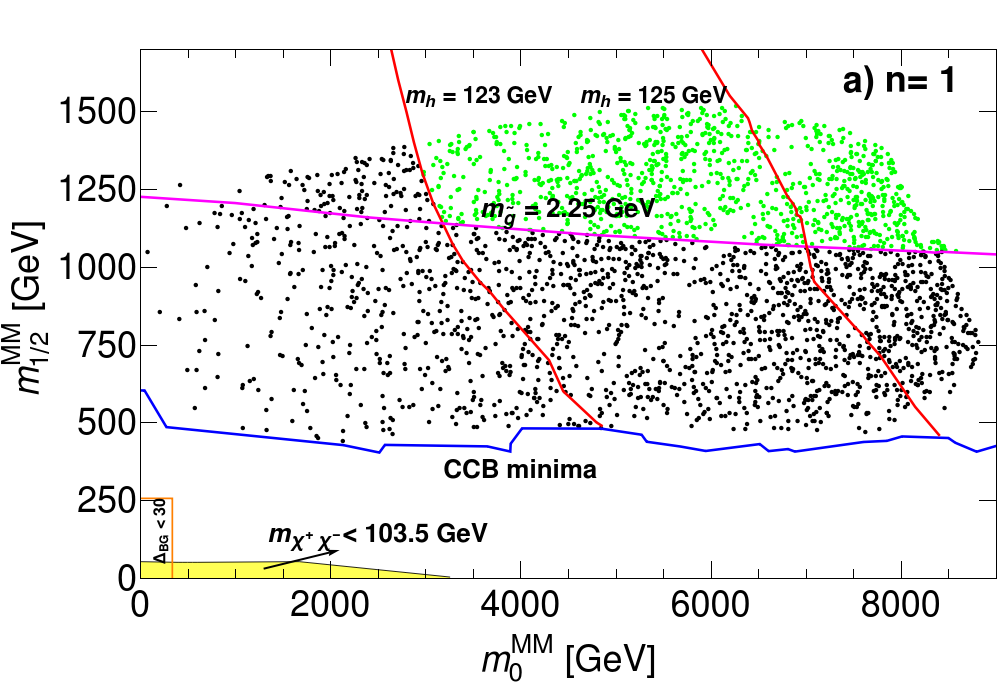}}\quad
  {\includegraphics[width=.48\textwidth]{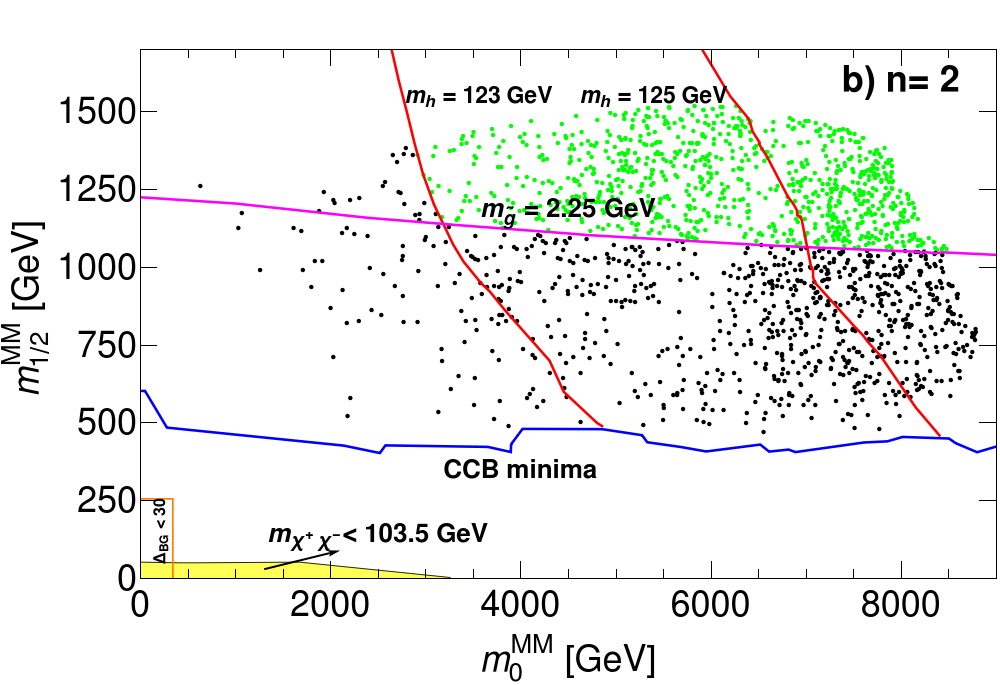}}\\ 
  {\includegraphics[width=.48\textwidth]{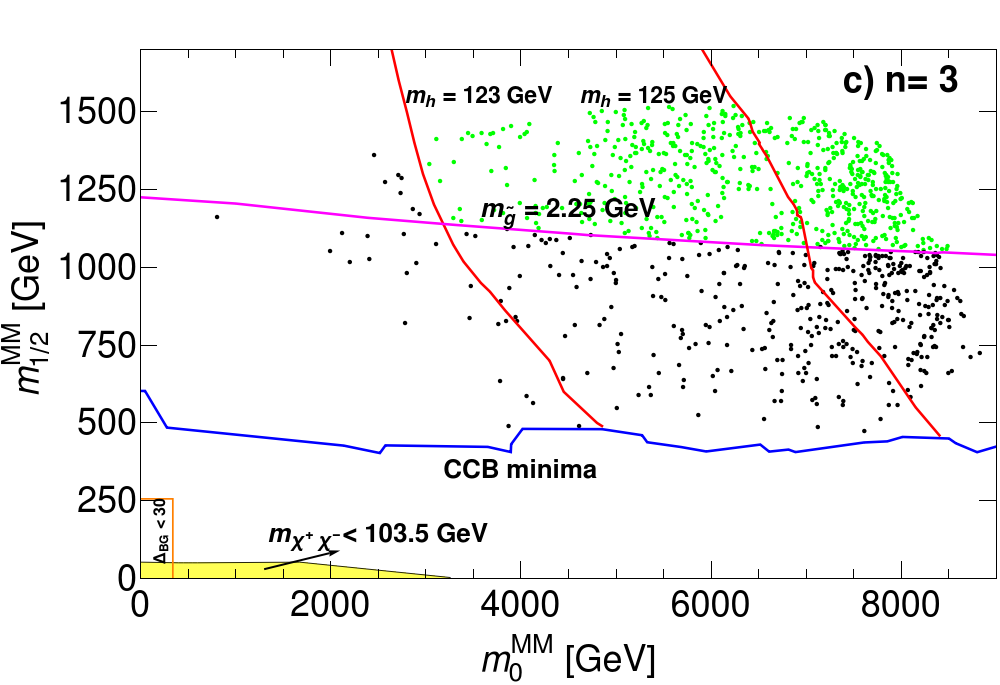}} \quad 
  {\includegraphics[width=.48\textwidth]{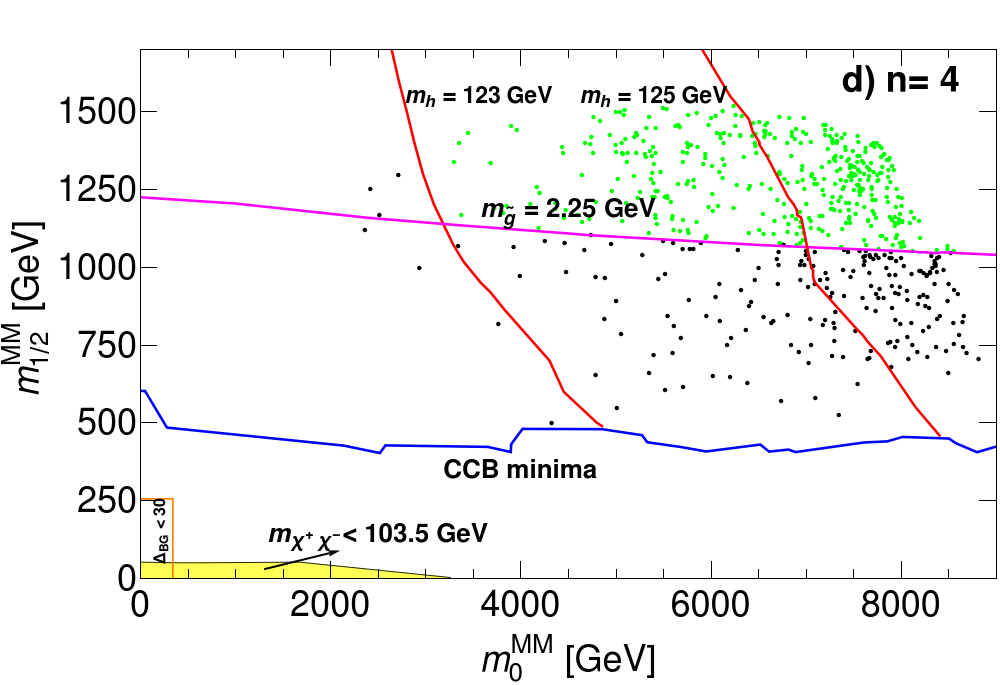}}\\
  \caption{For $m_{3/2}=20$ TeV, we plot 
the GMM parameter space in the $m_0^{MM}$ vs. $m_{1/2}^{MM}$ 
parameter space for $a_3=1.6\sqrt{c_{m}}$ with $c_{m3}=c_m$
and $\tan\beta =10$ with $m_A=2$ TeV. We plot for a 
landscape draw of {\it a}) $n=1$, {\it b}) $n=2$, {\it c}) $n=3$
and {\it d}) $n=4$ with $m_Z^{PU}<4 m_Z^{OU}$.
}  
\label{fig:m0mhf}
\end{figure}

In Fig. \ref{fig:m0mhf}{\it a}), we show the $m_0^{MM}$ vs. 
$m_{1/2}^{MM}$ plane for an $n=1$ landscape draw but with 
$a_3=1.6\sqrt{c_m}$, with $c_m=c_{m3}$ and with $\tan\beta =10$, 
$m_A=2$ TeV and $\mu =200$ GeV.
The lower-left yellow region shows where $m_{\tw_1}<103.5$ GeV in violation
of LEP2 constraints. Also, the lower-left orange box shows where
$\Delta_{BG}<30$ (old naturalness calculation). The bulk of the low
$m_{1/2}$ region here leads to tachyonic top-squark soft terms owing to 
the large trilinear terms $A_0^{MM}\equiv -a_3\alpha (m_{3/2}/16\pi^2)$. 
This region is nearly flat with increasing $m_0$ mainly because the larger
we make the GUT scale top-squark squared mass soft terms, 
the larger is the cancelling correction from RG running.
For larger $m_{1/2}$ values, then we obtain viable EW vacua since large
values of $M_3$ help to enhance top squark squared mass running to large
positive values (see {\it e.g.} Eq. 9.16h of Ref. \cite{WSS}).
The dots show the expected statistical result of scanning the landscape,
and the larger density of dots on the plot corresponds to greater
{\it stringy naturalness}. We also show the magenta contour of
$m_{\tg}=2.25$ TeV, below which is excluded by LHC gluino pair searches.
We also show contours of $m_h=123$ and 125 GeV. The green points are
consistent with LHC sparticle search limits and Higgs mass measurement.
From the plot, we see that the region of high stringy naturalness 
tends to lie safely beyond LHC sparticle search limits while at the same
time yielding a Higgs mass $m_h\simeq 125$ GeV.

In Fig's \ref{fig:m0mhf}{\it b}), {\it c}) and {\it d}), we increase
the power law statistical selection of soft terms to $n=2$, 3 and 4, 
respectively.\footnote{The relative density of dots between different
frames in Fig. \ref{fig:m0mhf} has no meaning.} 
As $n$ increases, then large soft terms are increasingly
favored until one hits the region for very large $m_{1/2}^{MM}$ 
and $m_0^{MM}$ where
contributions to the weak scale exceed a factor of 4 above our measured value.
The density of dots increasingly moves out towards large values of
$m_0^{MM}$ and $m_{1/2}^{MM}$ as $n$ increases. 
This is an example of {\it living dangerously} in the landscape
as noted by Arkani-Hamed, Dimopoulos and Kachru\cite{ArkaniHamed:2005yv}.
Then we see that the region beyond LHC gluino mass limits becomes
increasingly stringy natural! This is in sharp contrast to expectations from
conventional naturalness which favors sparticle masses close to the weak
scale\cite{Baer:2019cae}. 
For stringy naturalness, a value $m_{\tg}=3$ TeV is more natural
than a value of $m_{\tg}=300$ GeV! 
Thus, we see that the predictions from mirage-mediated landscape SUSY
are in close accord with what LHC is currently seeing: a Higgs mass
of $m_h\simeq 125$ GeV but as yet no sign of sparticles.

Finally, to compare and contrast the GMM model to the NUHM2 model with
universal gaugino masses, we list in Table \ref{tab:bm} two benchmark 
models computed using Isajet 7.88\cite{isajet}. Here, we have selected
a GMM$^{\prime}$ model with $\alpha$ chosen so that $m_{1/2}^{MM}=m_{1/2}=1250$ GeV,
$m_0^{MM}=m_0=5000$ GeV and $A_0^{MM}=A_0=-1.6 m_0=-8000$ GeV. 
Both cases contain $\tan\beta =10$, $\mu =200$ and $m_A=2$ TeV. 
The AMSB contribution to soft terms is fixed for GMM$^{\prime}$ by choosing
$m_{3/2}=20$ TeV. From Table \ref{tab:bm}, we see that the scalar mass 
spectrum is heavy and rather similar for the two cases. 
For the gaugino spectrum, we see that while 
$m_{\tw_2}\sim m_{\tz_4}\sim M_2\sim 1100$ GeV 
for both models, the gluino mass $m_{\tg}\sim 2556$ GeV for GMM$^{\prime}$ 
which is rather less than the value $m_{\tg}\sim 2931$ GeV for NUHM2.
Also, we see that $m_{\tz_3}\sim M_1\sim 748$ GeV for GMM$^{\prime}$ 
while $m_{\tz_3}\sim 562$ GeV for NUHM2. Thus, the gaugino masses are
{\it compressed} in GMM$^{\prime}$  compared to the gauginos from NUHM2
with a universal value of $m_{1/2}$ at $m_{GUT}$. 
Both models have a cluster of higgsinos around $\mu\sim 200$ GeV 
so these models may be difficult to distinguish at LHC upgrades. 
It may require an $e^+e^-$ collider operating with $\sqrt{s}>2m(higgsino)$ 
to measure the gaugino masses indirectly via their contribution 
to higgsino mass splitting. 
Such a collider could then distinguish mirage unification of gauginos
compared to GUT scale unified gaugino masses\cite{Baer:2014yta}.
\begin{table}\centering
\begin{tabular}{lcc}
\hline
parameter & $NUHM2$ & $GMM^\prime$\\
\hline
$m_0$      & 5000 & $\textendash$ \\
$m_{1/2}$        & 1250 & $\textendash$ \\
$A_0$      & -8000 & $\textendash$ \\
$\tan\beta$    & 10 & 10  \\
$m_{3/2}$      & $\textendash$ & 20000  \\
$\alpha$   & $\textendash$ & 9.9\\
$c_m$      & $\textendash$ & 16\\
$c_{m3}$    & $\textendash$ & 16\\
$a_3$      & $\textendash$ & 6.4\\
\hline
$\mu$          & 200   & 200 \\
$m_A$          & 2000  & 2000 \\
\hline
$m_{\tg}$   & 2931.4  & 2556.5 \\
$m_{\tu_L}$ & 5479.6 & 5305.3\\
$m_{\tu_R}$ & 5598.3 & 5432.8\\
$m_{\te_R}$ & 4822.6  & 4827.9\\
$m_{\tst_1}$ & 1750.2 & 1646.2\\
$m_{\tst_2}$ & 3953.6 & 3803.6\\
$m_{\tb_1}$ & 3987.4 & 3836.7 \\
$m_{\tb_2}$ & 5322.1 & 5169.5 \\
$m_{\ttau_1}$ & 4745.2 & 4752.2 \\
$m_{\ttau_2}$ & 5116.3 & 5094.0 \\
$m_{\tnu_{\tau}}$ & 5122.8 & 5101.0 \\
$m_{\tw_2}$ & -1061.2 & -1116.9 \\
$m_{\tw_1}$ & -210.0 & -210.1 \\
$m_{\tz_4}$ & -1074.7 & -1129.9\\ 
$m_{\tz_3}$ & -562.3 & -748.5 \\ 
$m_{\tz_2}$ & 208.2 & 207.8 \\ 
$m_{\tz_1}$ & -198.3 & -199.7 \\ 
$m_h$       & 124.8 & 124.2 \\ 
\hline
$\Omega_{\tz_1}^{std}h^2$ & 0.011 & 0.010  \\
$BF(b\to s\gamma)\times 10^4$ & 3.1 & 3.1 \\
$BF(B_s\to \mu^+\mu^-)\times 10^9$ & 3.8 & 3.8\\
$\sigma^{SI}(\tz_1, p)$ (pb) & $0.16\times10^{-8}$ & $0.11\times10^{-8}$ \\
$\sigma^{SD}(\tz_1 p)$ (pb) & $0.33\times10^{-4}$ &  $0.21\times10^{-4}$  \\
$\langle\sigma v\rangle |_{v\to 0}$  (cm$^3$/sec)  & $0.2\times10^{-24}$ &  
$0.2\times10^{-24}$ \\
$\Delta_{\rm EW}$ & 24.4 & 18.2\\
\hline
\end{tabular}
\caption{Input parameters and masses in~GeV units
for a natural mirage mediation SUSY benchmark point
as compared to a similar point from the NUHM2
model with $m_t=173.2$ GeV. The input parameters for 
the natural mirage mediation model such as $\alpha$ and $c_m$
have been calculated from $m_0^{MM}$ and $m_{1/2}^{MM}$
which are taken equal to $m_0$ and $m_{1/2}$ respectively as in NUHM2 model.
The $c_m$ and $c_{m3}$ have been taken equal to each other 
so that masses of first/second and third generation sfermions 
are equal at the GUT scale so as to match the NUHM2 model.
}
\label{tab:bm}
\end{table}

\section{Conclusions}
\label{sec:conclude}

From rather general considerations of the string landscape, it is to be 
expected that there is a statistical power law preference $m_{soft}^n$ for
soft SUSY breaking terms as large as possible, subject to the
anthropic condition that electroweak symmetry is properly broken and that
the pocket universe value of the weak scale does not exceed a factor 2-5
(here we use 4) from its measured value in our universe. Such a scenario is 
apt to lift the gravitino mass $m_{3/2}$ into the tens of TeV range such that
AMSB SSB terms are comparable to the weak scale. In such a case, then one 
expects moduli-mediated and anomaly-mediated soft terms to be
comparable and in such a setting the appropriate $N=1$ SUGRA framework 
is that of generalized mirage-mediation.

Within the GMM model and including a natural solution to the SUSY $\mu$ 
problem, we have made statistical predictions for model parameters
and sparticle and Higgs boson mass values for the cases of $n=1$ and 2 with
$m_{3/2}=20$ TeV.
For $n=1$ with $m_{3/2}=20$ TeV we find the mirage mediation scale 
$\mu_{mir}\sim 10^{10}-2\times 10^{14}$ GeV while for $n=2$ 
then $\mu_{mir}\sim 8\times 10^{12}-3\times 10^{14}$ GeV. 
These predictions can be somewhat falsified by measuring the gaugino masses
at LHC or a high energy $e^+e^-$ collider and extrapolating their masses via 
renormalization group running to find their intersection point $\mu_{mir}$, 
which then determines the mixing parameter $\alpha$. In this happy event,
then one could also {\it directly extract the gravitino mass $m_{3/2}$}. 
The mirage-mediation scenario would be rather implausible 
if no mirage mediation scale was found (the three gaugino masses did not unify
at a point) or if $\mu_{mir}$ was found to lie outside these ranges.

Regarding Higgs and sparticle mass predictions, the light Higgs boson mass 
is found to peak rather sharply around $m_h\simeq 125$ GeV. 
This is understood in part because the trilinear SSB term is pulled to large-- 
but not too large-- values such that there is large mixing in the stop sector
leading to large radiative corrections to $m_h$. 
The Higgs mass cannot get too large lest SUSY radiative corrections to the 
weak scale drive the value of the pocket-universe weak scale $m_Z^{PU}$ 
beyond the Agrawal {\it et al.}\cite{Agrawal:1997gf} 
anthropic window of allowed values. 

Meanwhile, the gluino is pulled up to $m_{\tg}\sim 3.5\pm 1.5$ TeV and the 
light top squark is pulled to $m_{\tst_1}\sim 1.5\pm 0.5$ TeV. 
With such large values of $m_{\tg}$ and $m_{\tst_1}$, an energy upgrade 
of LHC may be needed to realize SUSY discovery via gluino and/or top-squark
pair production. The pseudoscalar Higgs boson $m_A\sim 3.5\pm 1.5$ TeV so it
seems typically beyond the projected reach of LHC luminosity upgrades.
The most likely avenue for SUSY discovery at LHC would be via direct Higgsino
pair production $pp\to \tchi_1^0\tchi_2^0\to \ell^+\ell^-+\eslt $ where
the presence of an initial state jet radiation may help to trigger on
the expected soft dilepton signature\cite{SDLJMET}.
The soft dilepton invariant mass is expected to be bounded by 
$m_{\tz_2}-m_{\tz_1}\sim 5-10$ GeV. 
In fact, such a soft opposite-sign dilepton excess seems to be building 
in Atlas data.
Precision measurement of higgsino pair production also presents excellent 
motivation for construction of an $e^+e^-$ collider with 
$\sqrt{s}>2m(higgsino)\simeq 2\mu\simeq 400-600$ GeV\cite{Baer:2014yta}.

{\it Acknowledgements:} 

HB thanks the University of Colorado Department of Physics and Astrophysics
for hospitality while this work was completed. DS thanks the Fermilab Theory Group 
for hospitality when part of this work was done.
This work was supported in part by the US Department of Energy, Office
of High Energy Physics. 


%
\end{document}